\documentclass[journal]{IEEEtran}
\usepackage{acronym}
\usepackage{amsfonts}
\usepackage[dvips]{graphicx}
\usepackage{times}
\usepackage{cite}
\usepackage{amsmath}
\usepackage{array}
\usepackage{amssymb}
\usepackage{stfloats}
\usepackage{diagbox}
\usepackage{graphicx}
\usepackage{footnote}
\usepackage{amsthm}
\usepackage{booktabs}
\usepackage{array}
\usepackage[ruled,vlined]{algorithm2e}
\usepackage{subeqnarray}
\usepackage{cases}
\usepackage{threeparttable}
\usepackage{color}
\usepackage{epstopdf}
\usepackage{multirow}
\usepackage{tabularx}
\usepackage{enumerate}
\usepackage{multicol}
\usepackage{hyperref}
\usepackage{subfig}
\usepackage{caption}
\def\bb0{{\mathbb{0}}}

\def\bb{{\mathbf{b}}}

\def\bd{{\mathbf{d}}}
\def\bee{{\mathbf{e}}}

\def\bh{{\mathbf{h}}}

\def\bm{{\mathbf{m}}}
\def\bn{{\mathbf{n}}}

\def\bp{{\mathbf{p}}}
\def\bq{{\mathbf{q}}}
\def\br{{\mathbf{r}}}

\def\bx{{\mathbf{x}}}
\def\by{{\mathbf{y}}}
\def\bz{{\mathbf{z}}}
\def\b0{{\mathbf{0}}}

\def\bA{{\mathbf{A}}}
\def\bB{{\mathbf{B}}}
\def\bC{{\mathbf{C}}}

\def\bE{{\mathbf{E}}}
\def\bF{{\mathbf{F}}}

\def\bH{{\mathbf{H}}}
\def\bI{{\mathbf{I}}}

\def\bN{{\mathbf{N}}}

\def\bR{{\mathbf{R}}}

\def\bV{{\mathbf{V}}}
\def\bW{{\mathbf{W}}}
\def\bX{{\mathbf{X}}}
\def\bY{{\mathbf{Y}}}

\def\bbC{{\mathbb{C}}}

\def\sf0{{\mathsf{0}}}

\def\rmd{{\mathrm{d}}}

\def\rmp{{\mathrm{p}}}

\def\rm0{{\mathrm{0}}}

\def\Nr{{N_\mathrm{R}}}

\def\Np{{N_\mathrm{p}}}

\def\Nc{{N_\mathrm{c}}}

\def\Nt{{N_{\mathrm{t}}}}
\def\Nr{{N_{\mathrm{r}}}}

\def\Np{{N_{\mathrm{p}}}}
\def\Nd{{N_{\mathrm{d}}}}

\acrodef{CSI}[CSI]{channel state information}
\acrodef{CSIT}[CSIT]{channel state information at the transmitter}
\acrodef{CSIR}[CSIR]{channel state information at the receiver}
\acrodef{MIMO}[MIMO]{multiple-input multiple-output}
\acrodef{SISO}[SISO]{single-input single-output}
\acrodef{MISO}[MISO]{multiple-input single-output}
\acrodef{SIMO}[SIMO]{single-input multiple-output}
\acrodef{ADCs}[ADCs]{analog-to-digital convertors}
\acrodef{SNR}[SNR]{signal-to-noise ratio}
\acrodef{AWGN}[AWGN]{additive white Gaussian noise}
\acrodef{MRT}[MRT]{maximal ratio transmission}
\acrodef{DFT}[DFT]{Discrete Fourier Transform}
\acrodef{ULA}[ULA]{uniform linear array}
\acrodef{UPA}[UPA]{uniform planar array}
\acrodef{LS}[LS]{least squares}
\acrodef{ALMMSE}[ALMMSE]{approximate linear minimum mean squared error}
\acrodef{QIHT}[QIHT]{quantized iterative hard thresholding}
\acrodef{QIST}[QIST]{quantized iterative soft thresholding}
\acrodef{SVD}[SVD]{singular value decomposition}

\usepackage{bm}

\begin{document}

\title{Model-Driven Deep Learning for \\ MIMO  Detection}
\author{Hengtao He,~\IEEEmembership{Student Member,~IEEE,}
       Chao-Kai Wen,~\IEEEmembership{Member,~IEEE,} \\
       Shi Jin,~\IEEEmembership{Senior Member,~IEEE,}
       and Geoffrey Ye Li,~\IEEEmembership{Fellow,~IEEE}

\thanks{Manuscript received July 22, 2019; revised December 13, 2019 and February 10, 2020; accepted
Feb 23, 2020. Date of publication March XX, 2020; date of current version May XX, 2020. The work was supported in part by the National Key Research and Development Program 2018YFA0701602, the National Science Foundation of China (NSFC) for Distinguished Young Scholars with Grant 61625106, and the NSFC under Grant 61941104. The work of H. He was supported in part by the Scientific Research Foundation of Graduate School of Southeast University under Grant YBPY1939 and the Scholarship from the China Scholarship Council under Grant 201806090077. The work of C.-K. Wen was  supported  in  part  by  the  Ministry  of  Science  and  Technology  of  Taiwan  under  grants  MOST 108-2628-E-110-001-MY3  and  the  ITRI  in  Hsinchu,  Taiwan. The associate editor coordinating
the review of this paper and approving it for publication was Prof. Byonghyo Shim.
 \emph{(Corresponding author: Shi Jin.)}}

\thanks{This paper has been presented in part at IEEE Global Conference Signal and Information
Processing (Globalsip), Anaheim, CA, Nov. 2018 \cite{OAMP_Net}.}
\thanks{H.~He and S.~Jin are with the National Mobile Communications Research
Laboratory, Southeast University, Nanjing 210096, China (e-mail: hehengtao@seu.edu.cn, and jinshi@seu.edu.cn).}
\thanks{C.-K.~Wen is with the Institute of Communications Engineering, National
Sun Yat-sen University, Kaohsiung 804, Taiwan (e-mail: chaokai.wen@mail.nsysu.edu.tw).}
\thanks{G.~Y.~Li is with the School of Electrical and Computer Engineering,
Georgia Institute of Technology, Atlanta, GA 30332 USA (e-mail:
liye@ece.gatech.edu).}
}
\maketitle

\begin{abstract}

In this paper, we investigate the model-driven deep learning (DL) for MIMO detection. In particular, the MIMO  detector is specially designed by unfolding an iterative algorithm and adding some trainable parameters.  Since the number of trainable parameters is much fewer than the data-driven DL based signal detector, the model-driven DL based MIMO detector can be rapidly trained with a much smaller data set.  The proposed MIMO detector can be extended to soft-input soft-output detection easily. Furthermore, we investigate joint MIMO channel estimation and signal detection (JCESD), where the detector takes channel estimation error and channel statistics into consideration while channel estimation is refined by detected data and considers the detection error. Based on numerical results, the model-driven DL based MIMO detector significantly improves the performance of corresponding traditional iterative detector, outperforms other DL-based MIMO detectors and exhibits superior robustness to various mismatches.
\end{abstract}


\IEEEpeerreviewmaketitle

\section{Introduction}
Multiple-input multiple-output (MIMO) technology  can dramatically improve the spectral efficiency and link reliability and has been applied to many wireless communication systems. To obtain the benefits of MIMO \cite{diversity03Tse}, efficient channel estimation and signal detection algorithms, which balance performance and complexity, are essential in receiver design and have arouse a series of research \cite{AMP, MIMO_Detection, SD, AMPJSTSP, EPdetector}. Among existing detectors, maximum likelihood (ML) detection can achieve the optimal performance. However,  its complexity increases exponentially with the number of decision variables. Some suboptimal linear detectors, such as zero-forcing (ZF) and linear minimum mean-squared error (LMMSE) detectors, are with reduced computational complexity, but have a huge performance degradation compared with the ML detection.

With excellent performance and moderate complexity, iterative detectors, based on approximate message passing (AMP) \cite{Donohol_AMP} and expectation propagation (EP) \cite{EP} algorithms, have been proposed for  MIMO detection \cite{AMPJSTSP,EPdetector}. The AMP-based detector approximates the posterior probability on a dense factor graph by using the central limit theorem and the Taylor expansion, which can achieve Bayes-optimal performance in the large-scale systems when the elements of the channel matrix are with independent and identically sub-Gaussian distribution. The EP-based detector \cite{EPdetector,EP} is derived by approximating the posterior distribution with  factorized Gaussian distributions and can achieve Bayes-optimal performance when the channel matrix is unitarily-invariant \footnote{A matrix $\mathbf{A}=\mathbf{U}\mathbf{\Sigma}\mathbf{V}$ is unitarily-invariant if $\mathbf{U}$, $\mathbf{\Sigma}$ and $\mathbf{V}$ are mutually independent, and $\mathbf{U}$, $\mathbf{V}$ are Haar-distributed. The independent and identically distributed (i.i.d.) Gaussian matrix is a typical unitarily-invariant matrix.} and with a large scale. However, for practical small-size (e.g., $4 \times 4$ or $8 \times 8$) MIMO systems, the performance of these iterative detectors is still far from Bayes-optimal solution and has serious deterioration with correlated MIMO channels and imperfect channel state information (CSI) \cite{EP17csi}. 

Owing to strong learning ability from the data, deep learning (DL) has been successfully introduced to computer vision, automatic speech recognition, and natural language processing. Recently, it has been applied in physical layer communications \cite{DL2017wang, Modeldriven18DL, DL2018Qin}, such as channel estimation \cite{DL2018HE,CE2019Yang,DLdata_adided}, CSI feedback \cite{DL2018Wen}, signal detection \cite{DL2OFDM,DeepMIMO,Improving,CHemp18Tan,DL18SD,TPG18,GDNet,MMNet,ComNet18}, channel coding \cite{DL18coding, TurboNet}, and end-to-end transceiver design \cite{OShea,GAN}. In particular, a five-layer fully connected deep neural network (DNN) is embedded into an orthogonal frequency-division multiplexing (OFDM) system for joint channel estimation and signal detection (JCESD) by treating the receiver as a black box and without exploiting domain knowledge \cite{DL2OFDM}. However, training such a black-box-based network requires a lot of training time in addition to a huge data set. On the other hand, model-driven DL constructs the network topology based on known  domain knowledge and has been successfully applied to image reconstruction \cite{ADMM_Net}, sparse signal recovery \cite{AMP_Net,TISTA,ALISTA,ICML}, and  wireless communications recently\cite{OAMP_Net,Modeldriven18DL}.

For MIMO detection, a specifically designed network, named DetNet, has been proposed in \cite{DeepMIMO} by unfolding the iteration of a projected gradient descent algorithm and adding considerable trainable variables. DetNet has comparable performance with the AMP-based detector and is more robust to ill-conditioned channels \cite{DeepMIMO}. To further reduce the number of learnable parameters and improve convergence, the approaches in \cite{Improving} and \cite{CHemp18Tan} use DL techniques for the belief propagation and message passing detector, respectively. In \cite{DL18SD},  a DL-based sphere
decoding algorithm is proposed, where the radius of the decoding hypersphere is learned by  DNN. The performance achieved by this algorithm is very close to the optimal ML detection. However, most of the existing DL-based detector assume accurate CSI at the receiver and ignore the channel estimation error.

Motivated by existing works, we develop a model-driven DL network, named OAMP-Net2, for MIMO detection in this article, where the iterative detector is improved with a few number of trainable variables to adapt to various channel environments. The structure of the detector is obtained by unfolding the OAMP detector, which is similar to our early work in \cite{OAMP_Net} and inspired by the TISTA network\cite{TISTA} but adds more trainable parameters. Furthermore, an OAMP-Net2-based JCESD architecture is proposed for imperfect CSI and data-aided scheme is utilized to further improve channel estimation. The trainable parameters are optimized by DL technique to adapt to various channel environments and take channel estimation error into consideration. The main contributions of this paper are summarized as follows:

\begin{itemize}

  \item Different from the existing DL-based MIMO detector \cite{DeepMIMO,Improving,TPG18,CHemp18Tan,GDNet,MMNet,DL18SD} with perfect CSI, we consider the MIMO detection with estimated channel, which improves the performance of MIMO receiver by considering the characteristics of channel estimation error and channel statistics and using the estimated payload data to refine the channel estimation.

  \item Compared with the existing DL-based MIMO detector \cite{DeepMIMO,Improving,CHemp18Tan,DL18SD}, our proposed detector can provide soft-output information for decoder and absorb the soft information. In addition, only a few trainable parameters are required to be learned, which can reduce the demand for computing resources and training time significantly.

  \item Based on our numerical results, the OAMP-Net2 has considerable performance gain compared with the OAMP detector. Furthermore, OAMP-Net2 has strong robustness to signal-to-noise (SNR), channel correlation, modulation symbol and MIMO configuration mismatches.
\end{itemize}

\emph{Notations}---For any matrix $\bA$, $\bA^{T}$, $\bA^{H}$, and ${ \mathrm{tr}}(\bA)$ denote the transpose,  conjugated transpose, and  trace of $\bA$, respectively. In addition, $\mathbf{I}$ is the identity matrix, $\mathbf{0}$ is the zero matrix. A proper complex Gaussian with mean $\boldsymbol{\mu}$ and covariance $\boldsymbol{\Omega}$ can be described by the probability density function,
\begin{equation*}
  \mathcal{N}_{\mathbb{C}}(\mathbf{z};\boldsymbol{\mu},\boldsymbol{\Theta})=\frac{1}{\mathrm{det}(\pi \boldsymbol{\Theta})}
  e^{-(\mathbf{z}-\boldsymbol{\mu})^{H}\boldsymbol{\Theta}^{-1}(\mathbf{z}-\boldsymbol{\mu})}.
\end{equation*}

The rest of this paper is organized as follows. After introducing the JCESD architecture  in Section \ref{JCD_section}, we  develop channel estimator in Section \ref{CE} and propose the model-driven DL  detector in Section \ref{detector}.
Then, numerical results are presented in Section \ref{Simulation}.
Finally, Section \ref{con} concludes the paper.

\section{Joint Channel Estimation and Signal Detection}\label{JCD_section}
In this section,  we consider a MIMO system with $\Nt$ transmit and $\Nr$ receive antennas. After presenting the JCESD architecture\footnote{One can introduce various JCESD architectures for MIMO systems, including the schemes in \cite{Takeuchi,Data_aid,BayesJCD,Turbo16like,JCESD}, we present a turbo-like JCESD architecture similar to\cite{Turbo16like} in this paper.}, we introduce the signal detection considering channel estimation error and data-aided channel estimation\footnote{Although we mainly investigate the model-driven-DL-based MIMO detector in this paper, we first introduce a JCESD architecture and then elaborate the channel estimator and signal detector modules, respectively.}.

We assume that  channel matrix $\bH \in \bbC^{\Nr \times \Nt}$ does not change in a time slot. In each time slot, $\Np $ pilot vectors $\bx_{\rmp}[n] \in \bbC^{\Nt \times 1} $ for $n=1,\ldots,\Np$, are first transmitted, which are followed $N_{\rmd}$ data vectors, $\bx_{\rmd}[n] \in \bbC^{\Nt \times 1} $.

The received signal vectors are $\by_{\rmp}[n] \in \bbC^{ \Nr \times 1} $ for $n=1,\ldots,\Np$ and $\by_{\rmd}[n] \in \bbC^{\Nr \times 1}$ for $n=1,\ldots,\Nd$ corresponding to the pilot and data vectors, respectively. We can also express them into matrix forms as $\bX_{\rmp} = \left(\bx_{\rmp}[1],\ldots, \bx_{\rmp}[\Np] \right) \in \bbC^{\Nt \times \Np} $, $\bY_{\rmp} = \left(\by_{\rmp}[1],\ldots, \by_{\rmp}[\Np] \right) \in \bbC^{\Nr \times \Np} $, $\bX_{\rmd} = \left(\bx_{\rmd}[1],\ldots, \bx_{\rmd}[\Nd] \right) \in \bbC^{\Nt \times \Nd}$, and $\bY_{\rmd} = \left(\by_{\rmd}[1],\ldots, \by_{\rmd}[\Nd] \right)\in \bbC^{\Nr \times \Nd}$.
\subsection{JCESD Architecture}\label{JCESD}
As in Fig.\,$1$, we consider a turbo-like JCESD architecture for MIMO systems in this paper, which shares the same spirit as iterative decoding. In JCESD, channel estimator and signal detector exchange information iteratively until convergence \cite{Turbo16like}. In the first iteration, pilot-only based channel estimation is performed. In the subsequent iterations, data-aided channel estimation is employed with the help of the detected data.
\begin{figure*}
  \centering
  \includegraphics[width=16cm]{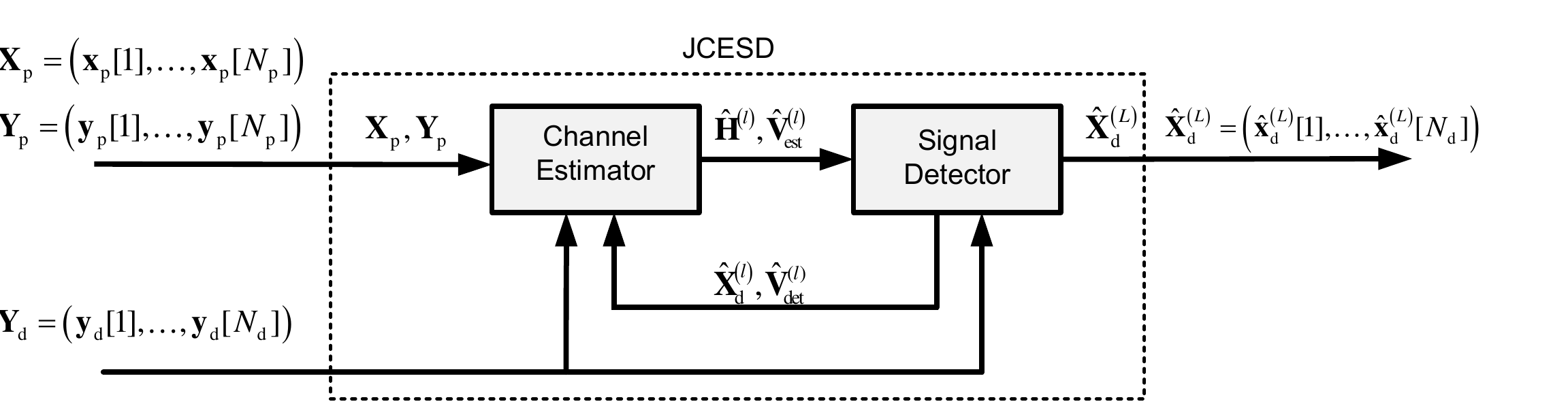}
  \caption{.~~The diagram of the turbo-like JCESD architecture. The channel estimator and signal detector exchange information iteratively until convergence.}\label{JCD}
\end{figure*}

The input of the JCESD architecture is the pilot signal matrix, $\mathbf{X}_{\rmp}$, received signal matrix corresponding to the pilot matrix, $\bY_{\rmp}$, corresponding to the data matrix, $\bY_{\rmd}$ in each time slot. In the $l$-th turbo iteration, $\hat{\bH}^{(l)}$ is the estimated channel matrix, $\hat{\bX}_{\rmd}^{(l)}$ is the estimated data matrix, and $\hat{\bV} _{\mathrm{est}}^{(l)}$ and $\hat{\bV}^{(l)}_{\mathrm{det}}$ are used to compute the covariance matrix for equivalent noise in signal detector and channel estimator, respectively. The final output of the signal detector is finally detected data matrix $\hat{\bX}_{\mathrm{d}}^{(L)}$, where $L$ is the total number of turbo iterations.

 Compared with the conventional receiver design where the channel estimator and signal detector are designed separately, this architecture can improve the performance of the receiver by considering the characteristics of channel estimation error in addition to channel statistics when performing signal detection and using the estimated payload data for channel estimation, as we will illustrate subsequently.
\subsection{Signal Detection with Channel Estimation Error}\label{SD_CE}
In the MIMO system, the received data signal vector $\by_{\rmd}[n]$ corresponding to the $n$-th data vector can be expressed by
\begin{equation}\label{eq1}
  \by_{\rmd}[n]=\bH\bx_{\rmd}[n]+\bn_{\rmd}[n],
\end{equation}
where $\bn_{\rmd}[n] \sim \mathcal{N}_{\mathbb{C}}(0,\sigma^{2}\mathbf{I}_\Nr) $  is the additive white Gaussian noise (AWGN) vector. Note that (\ref{eq1}) can also be expressed into matrix from as
\begin{equation}\label{eqSD1}
  \bY_{\rmd}=\bH\bX_{\rmd}+\bN_{\rmd},
\end{equation}
where $\bN_{\rmd} = \left(\bn_{\rmd}[1],\ldots, \bn_{\rmd}[\Nd] \right)\in \bbC^{\Nr \times \Nd} $ is the AWGN matrix in the data transmission stage.
Denote the estimated channel $\hat{\bH} \in \mathbb{C}^{\Nr\times \Nt}$ as
\begin{equation}\label{channel_error}
  \hat{\mathbf{H}} = \mathbf{H} + \bm{\Delta}\bH,
\end{equation}
where $\bm{\Delta}\bH$ is the channel estimation error. If the estimated channel is used for signal detector, the signal detection problem can be formulated as
\begin{align}\label{eqSD_CSI}
  \by_{\rmd}[n] & =\bH\bx_{\rmd}[n]+\bn_{\rmd}[n] \nonumber \\
 & = (\hat{\bH} - \bm{\Delta}\bH)\bx_{\rmd}[n] + \bn_{\rmd}[n] \nonumber \\
 & = \hat{\bH}\bx_{\rmd}[n] + \bn_{\rmd}[n]-\bm{\Delta}\bH\bx_{\rmd}[n] \nonumber \\
 & = \hat{\bH}\bx_{\rmd}[n] + \hat{\bn}_{\rmd}[n],
\end{align}
where $\hat{\bn}_{\rmd}[n] = \bn_{\rmd}[n] - \bm{\Delta}\bH\bx_{\rmd}[n]$ is the equivalent noise in signal detector which includes the contribution of  channel estimation error and original additive noise.  $\hat{\bn}_{\rmd}[n] \sim \mathcal{N}_{\mathbb{C}}(\mathbf{0},\hat{\bV}_{\rmd}[n])$ is assumed to be Gaussian distribution\footnote{The assumption that $\hat{\bn}_{\rmd}[n]$ is Gaussian distribution is reasonable as we consider the LMMSE channel estimator in next section.}, and the covariance matrix $\hat{\bV}_{\rmd}[n]$ can be obtained by considering the statistical properties of the channel estimation error and detailed calculated process is shown in Appendix A.
\subsection{Data-Aided Channel Estimation}\label{Data_aided}
In this section, we introduce the  data-aided channel estimation approach. Conventional pilot-only based channel estimation is performed and transmitted symbols are detected in the first iteration. Then, the detected symbols are fed back to the channel estimator as additional pilot symbols to refine the channel estimation.

In the channel training stage, pilot matrix $\bX_{\rmp}$ is transmitted, similar to (\ref{eq1}), the received signal matrix corresponding to the pilot matrix $\bY_{\rmp}\in\mathbb{C}^{\Nr\times \Np}$  can be expressed as
\begin{equation}\label{eq_CE}
  \bY_{\rmp}=\bH\bX_{\rmp}+\bN_{\rmp},
\end{equation}
where $\bN_{\rmp} = \left(\bn_{\rmp}[1],\ldots,\mathbf{n}_{\rmp}[\Np]\right)  \in \mathbb{C}^{\Nr\times \Np}$  is the AWGN matrix and each column $ \mathbf{n}_{\rmp}[n] \sim \mathcal{N}_{\mathbb{C}}(0,\sigma^{2}\mathbf{I}_{N_{r}}) $ for $n = 1,\ldots, \Np $. 
The estimated data matrix $\hat{\mathbf{X}}_{\rmd}$ can be expressed as
\begin{equation}\label{channel_error}
 \hat{\bX}_{\rmd} = \bX_{\rmd} + \bE_{\rmd},
\end{equation}
where $\mathbf{E}_{\rmd}$ is the signal detection error matrix. In data-aided channel estimation stage, estimated $\hat{\bX}_{\rmd}$ are fed back to channel estimator as additional pilot. Then, the received signal matrix $\bY_{\rmd}$ corresponding to $\hat{\bX}_{\rmd}$ can be expressed as
\begin{align}\label{eqCE_data1}
\bY_{\rmd} & = \bH\bX_{\rmd}+\bN_{\rmd} \nonumber \\
  &= \bH(\hat{\bX}_{\rmd}-\bE_{\rmd})+\bN_{\rmd} \nonumber \\
  &=  \bH \hat{\bX}_{\rmd} + (\bN_{\rmd}-\bH\bE_{\rmd})\nonumber \\
  &=  \bH \hat{\bX}_{\rmd}+\hat{\bN}_{\rmp},
\end{align}
where $\hat{\bN}_{\rmp} = \bN_{\rmd} - \bH\bE_{\rmd}$  is the equivalent noise for additional pilot part $\hat{\bX}_{\rmd}$. The statistical information of the $n$-th column of $\hat{\bN}_{\rmp}$,  $\hat{\mathbf{n}}_{\rmp}[n]\sim \mathcal{N}_{\bbC}(\mathbf{0},\hat{\mathbf{V}}_{\rmp}[n])$ for $n=1,\ldots,\Nd $, where $\hat{\mathbf{V}}_{\rmp}[n]$ is calculated in Appendix A and will be utilized in data-aided channel estimation stage. Then, we  denote $\bY = (\bY_{\rmp}, \bY_{\rmd})$ as received signal matrix corresponding to overall transmitted signal matrix. Based on (\ref{eq_CE}) and (\ref{eqCE_data1}), we have
\begin{align}\label{eqCE_data}
\bY & = \left(\bY_{\rmp} \ \bY_{\rmd}\right)  \nonumber \\
& = (\bH\bX_{\rmp}+\bN_{\rmp}, \bH \hat{\bX}_{\rmd} + \hat{\bN}_{\rmp})   \nonumber \\
& = \bH(\bX_{\rmp},  \hat{\bX}_{\rmd} )+ (\bN_{\rmp}, \hat{\bN}_{\rmp}) \nonumber \\
& = \bH\bX+\bN,
\end{align}
where $\bX = (\bX_{\rmp},  \hat{\bX}_{\rmd})$, $\bN = (\bN_{\rmp}, \hat{\bN}_{\rmp})$
can be interpreted as  the equivalent pilot signal and noise in data-aided channel estimation stage, respectively.
\subsection{Model-Driven DL for JCESD}
In Section \ref{JCESD}, we have introduced the principle of the JCESD architecture.  Compared with other DL-based JCESD architecture \cite{DL2OFDM,DLdata_adided}, which uses a large number of data to train the black-box-based network, we construct the network architecture by employing model-driven DL and domain knowledge. In particular, we employ the LMMSE channel estimator and construct the  model-driven DL detector. As in \cite{OAMP_Net}, the proposed model-driven DL detector is obtained by unfolding the existing iterative detector and adding several trainable parameters, which fully exploiting the domain knowledge. This architecture is promising as it inherits the superiority of traditional approach and uses  DL technique to improve the performance. We will introduce the channel estimator and signal detector in detail in Sections \ref{CE} and \ref{detector}, respectively.
\section{LMMSE Channel Estimator}\label{CE}
To facilitate the representation of the channel estimation problem, we apply matrix vectorization to (\ref{eq_CE}) and rewrite it as
\begin{equation}\label{eq_vec_ce}
\by_{\rmp} = \bA_{\rmp} \mathbf{h} + \bn_{\rmp},
\end{equation}
where $\bA_{\rmp} = \bX_{\rmp}^{T} \otimes \bI_{\Nr} \in \mathbb{C}^{\Np\Nr\times \Nt \Nr}$,  $\by_{\rmp}=\mathrm{vec}(\bY_{\rmp}) \in \mathbb{C}^{\Np\Nr\times 1}$, $\bh=\mathrm{vec}(\mathbf{H}) \in \mathbb{C}^{\Nr\Nt \times 1} $ and $ \bn_{\rmp} = \mathrm{vec}(\mathbf{N}_{\rmp})\in \mathbb{C}^{\Np\Nr\times 1}$.  We denote $\otimes$ as the matrix Kronecker product and $\mathrm{vec}(\cdot)$ as the vectorization operation. In pilot-only based channel estimation stage, the LMMSE estimate of $\bh$ is given by
\begin{equation}\label{LMMSE}
 \hat{\bh}_{\rmp} = \bR_{\bh \bh}\bA_{\rmp}^{H}(\bA_{\rmp} \bR_{\bh \bh} \bA_{\rmp}^{H}+\sigma^{2}\bI_{\Np\Nr})^{-1}\by_{\rmp},
\end{equation}
where $\bR_{\bh \bh}$ is the channel covariance matrix. Based on the property of LMMSE estimate, $\hat{\bh}_{\rmp}$ is a Gaussian random vector. The channel estimation error vector $\bm{\Delta}{\bh_{\rmp}}=\hat{\bh}_{\rmp}-\bh$ is also a Gaussian random vector with zero-mean and the covariance matrix $ \bR_{\bm{\Delta}{\bh_{\rmp}}}$  can be computed as
\begin{equation}\label{LMMSE_error}
 \bR_{\bm{\Delta}{\bh_{\rmp}}} = \bR_{\bh \bh}-\bR_{\bh \bh}\bA_{\rmp}^{H}(\bA_{\rmp} \bR_{\bh \bh} \bA_{\rmp}^{H}+\sigma^{2}\bI_{\Np\Nr})^{-1}\bA_{\rmp}\bR_{\bh \bh}.
\end{equation}

In data-aided channel estimation stage, considering data feedback to channel estimator, the LMMSE channel estimation is given by
\begin{equation}\label{LMMSE2}
 \hat{\bh} = \bR_{\bh \bh}\bA^{H}(\bA \bR_{\bh \bh} \bA^{H}+\bR_{\bn \bn})^{-1}\by.
\end{equation}
where  $\bA = \bX^{T} \otimes \mathbf{I}_{\Nr} \in \mathbb{C}^{\Nc\Nr\times \Nt \Nr}$,  $\by=\mathrm{vec}(\mathbf{Y}) \in \mathbb{C}^{\Nc\Nr\times 1}$,  $ \bn= \mathrm{vec}(\mathbf{N}) \in \mathbb{C}^{\Nc\Nr \times 1} $ and $\Nc = \Np + \Nd$. The covariance matrix
$ \bR_{\bm{\Delta}\bh}$  of the  channel estimation error vector $\bm{\Delta}{\bh}=\hat{\bh}-\bh$ can be computed as
\begin{equation}\label{LMMSE_error2}
 \bR_{\bm{\Delta}\bh} = \bR_{\bh \bh}-\bR_{\bh \bh}\bA^{H}(\bA \bR_{\bh \bh} \bA^{H}+\bR_{\bn \bn})^{-1}\bA\bR_{\bh \bh}.
\end{equation}
The covariance matrix $\bR_{\bn \bn}$ contains the equivalent noise power of the actual pilot $\mathbf{X}_{\rmp}$ and additional pilot part $ \hat{\mathbf{X}}_{\rmd}$, which is calculated in Appendix A.
\section{Model-Driven DL Detector}\label{detector}
In this section, we develop a model-driven DL detector, called OAMP-Net2, which has been partly presented in our early work \cite{OAMP_Net}. After introducing the principle of the OAMP detector, we present the structure of OAMP-Net2 in detail. Afterwards, computational complexity  and extension to soft-input and soft-output detection are also investigated.
\subsection{OAMP Detector}
The OAMP algorithm has been proposed to solve sparse linear inverse problems in compressed sensing \cite{OAMP} and can be utilized for MIMO detection in Algorithm 1. The goal of the OAMP algorithm is to recover the transmitted signal vector $\bx_{\rmd}$ from the received signal vector $\by_{\rmd}=\bH\bx_{\rmd}+\bn_{\rmd}$\footnote{ As each transmitted symbol vector $\bx_\rmd(n)$ for $n=1,\ldots,\Nd $ in each time slot shares the same channel $\bH$, we omit the time index $n$ and use
$\bx_{\rmd}$ to refer $\bx_{\rmd}[n]$ for simplicity.}. The principle of the algorithm is to decouple the posterior probability $\mathcal{P}(\mathbf{x}_{\rmd}|\mathbf{y}_{\rmd},\hat{\mathbf{H}})$ into a series of
$\mathcal{P}(x_{i}|\mathbf{y}_{\rmd},\hat{\mathbf{H}}) (i=1,2,\ldots,\Nt)$ in an iterative way, when estimated channel $\hat{\bH}$ and received signal $\by_{\rmd}$ are available.
\begin{algorithm}\label{algGE}
\caption{OAMP algorithm for MIMO detection} 
\hspace*{0.02in} {\bf Input:} 
Received signal $\mathbf{y}_{\rmd}$, estimated channel matrix $\hat{\mathbf{H}}$, equivalent noise covariance matrix $\bR_{\hat{\bn}_{\rmd}\hat{\bn}_{\rmd}}$. \\
\hspace*{0.02in} {\bf Output:} 
Recovered signal $\hat{\mathbf{x}}_{\rmd, T+1}$.\\
\hspace*{0.02in} {\bf Initialize:}
$\tau_{1} \leftarrow 1$, $\hat{\mathbf{x}}_{\rmd,1}\leftarrow \mathbf{0}$
\begin{equation}\label{eqr}
  \mathbf{r}_{t}=\hat{\mathbf{x}}_{\rmd,t}+\mathbf{W}_{t}(\mathbf{y}_{\rmd}-\hat{\mathbf{H}}\hat{\mathbf{x}}_{\rmd,t}),
\end{equation}
\begin{equation}\label{eqs}
  \hat{\mathbf{x}}_{\rmd,t+1}=\mathtt{E}\left\{\mathbf{x}|\mathbf{\mathbf{r}}_{t},\tau_{t}\right\},
\end{equation}
\begin{equation}\label{eqv}
  v_{t}^{2}=\frac{\|\mathbf{y}-\hat{\mathbf{H}}\hat{\mathbf{x}}_{\rmd,t}\|_{2}^{2}- \mathrm{tr}(\bR_{\hat{\bn}_{\rmd}\hat{\bn}_{\rmd}})}{\mathrm{tr}(\hat{\mathbf{H}}^{H}\hat{\mathbf{H}})},
\end{equation}
\begin{equation}\label{eqt}
  \tau^{2}_{t}=\frac{1}{\Nt}\mathrm{tr}(\mathbf{B}_{t}\mathbf{B}_{t}^{H})v_{t}^{2}+\frac{1}{\Nt}\mathrm{tr}(\mathbf{W}_{t}\bR_{\hat{\bn}_{\rmd}\hat{\bn}_{\rmd}}\mathbf{W}_{t}^{H}).
\end{equation}
\end{algorithm}

The OAMP detector is presented in Algorithm 1 and mainly contains two modules: linear estimator (\ref{eqr}) and nonlinear estimator (\ref{eqs}). The error variance estimators (\ref{eqv}) and (\ref{eqt}) are the average variance of the two error vectors $\bp_{t}$ and $\bq_{t}$, where $\bp_{t}=\br_{t}-\bx_{\rmd}$ and $\bq_{t}=\hat{\bx}_{\rmd, t}-\bx_{\rmd}$ are used to measure the accuracy of the output in the linear and nonlinear estimators, respectively. They are defined as
\begin{equation}\label{eqvbar}
v_{t}^{2}=\frac{\mathtt{E}[\|\mathbf{q}_{t}\|^{2}_{2}]}{\Nt}, \, \tau_{t}^{2}=\frac{\mathtt{E}[\|\mathbf{p}_{t}\|^{2}_{2}]}{\Nt},
\end{equation}
and can be computed by employing (\ref{eqv}) and (\ref{eqt}).
%
%
\subsubsection{Linear Estimator}
The matrix $\mathbf{W}_{t}$ in linear estimator (\ref{eqr}) can be the transpose of $\hat{\mathbf{H}}$, the pseudo inverse of $\hat{\mathbf{H}}$ or the LMMSE matrix. From \cite{OAMP}, the optimal one is
\begin{equation}\label{UMMSE}
  \mathbf{W}_{t}=\frac{\Nt}{\mathrm{tr}(\hat{\mathbf{W}_{t}}\hat{\mathbf{H}})}\hat{\mathbf{W}_{t}},
\end{equation}
where $\hat{\mathbf{W}_{t}}$ is the LMMSE matrix given by
\begin{equation}\label{BMMSE}
 \hat{\mathbf{W}_{t}} = v_{t}^2\hat{\mathbf{H}}^{H}(v_{t}^2\hat{\mathbf{H}}\hat{\mathbf{H}}^{H}+\bR_{\hat{\bn}_{\rmd}\hat{\bn}_{\rmd}})^{-1},
\end{equation}
where $v_{t}^{2}$ is expressed in (\ref{eqv}), and $\bR_{\hat{\bn}_{\rmd}\hat{\bn}_{\rmd}}$ is the covariance matrix of the equivalent noise $\hat{\bn}_{\rmd}$ in signal detector, which includes the contribution of the channel estimation error $\bm{\Delta}\bH$ and original additive noise $\bn_{\rmd}$, and has been discussed in Section \ref{SD_CE}. The matrix $\bW_{t}$ is called de-correlated when $\mathrm{tr}(\bB_{t})=0$, where $\bB_{t}=\bI-\bW_{t}\hat{\bH}$, and thus ensures the entries of $\bp_{t}$ are uncorrelated with those of $\bx_{\rmd}$ and mutually uncorrelated with zero-mean and identical variances. 
\subsubsection{Nonlinear Estimator}
The nonlinear estimator in the OAMP detector is constructed by MMSE estimate of $\mathbf{x}_{\rmd}$, which is with respect to the equivalent AWGN channel
  \begin{equation}\label{eqAWGN}
    \mathbf{r}_{t}=\mathbf{x}_{\rmd,t}+ \mathbf{w}_{t},
  \end{equation}
  where $\mathbf{w}_{t}\sim \mathcal{N}_{\mathbb{C}}(\mathbf{0},\tau^{2}_{t}\mathbf{I})$. As the transmitted symbol $\mathbf{x}_{\rmd}$ is from the discrete constellation set $\mathcal{S}=\{s_{1},s_{2},\ldots,s_{M}\}$, corresponding MMSE estimate for each element of the estimated symbol vector is given by
\begin{equation}\label{eqE}
\hat{\mathbf{x}}_{\rmd,t+1}^{(i)} = \mathtt{E}\left\{x_{i}|r_{i},\tau_{t}\right\}=\frac{\sum_{s_{i}}s_{i}\mathcal{N}_{\mathbb{C}}(s_{i};r_{i}, \tau^{2}_{t})p(s_{i})}{\sum_{s_{i}}\mathcal{N}_{\mathbb{C}}(s_{i};r_{i}, \tau^{2}_{t})p(s_{i})},
\end{equation}
where $p(s_{i})$ is the prior distribution of the transmitted symbol $x_{i}$ and is given by
\begin{equation}\label{eqprior}
p(x_{i})=\sum_{j \in M}\frac{1}{\sqrt{M}}\delta(x_{i}-s_{j}).
\end{equation}

From (\ref{eqr}), (\ref{eqs}), (\ref{eqt}) and (\ref{eqAWGN}), we observe that $\mathbf{r}_{t}$ and $\tau^{2}_{t}$ are the prior mean and variance in MMSE estimator $(\ref{eqs})$ that control the accuracy  and convergence of estimated result $\hat{\mathbf{x}}_{\rmd, t+1}$. The OAMP detector uses an iterative manner to obtain $\mathbf{r}_{t}$ and $\tau^{2}_{t}$, and the step-size for the update of $\mathbf{r}_{t}$ and $\tau^{2}_{t}$ will influence the final performance.
Instead of using complicated analytical method to find an optimal step-size,  we use a DL approach for providing an appropriate step-size to update $\mathbf{r}_{t}$ and $\tau^{2}_{t}$ and constructing learnable nonlinear estimator to improve the detection performance.
\begin{figure*}
  \centering
  \includegraphics[width=16cm]{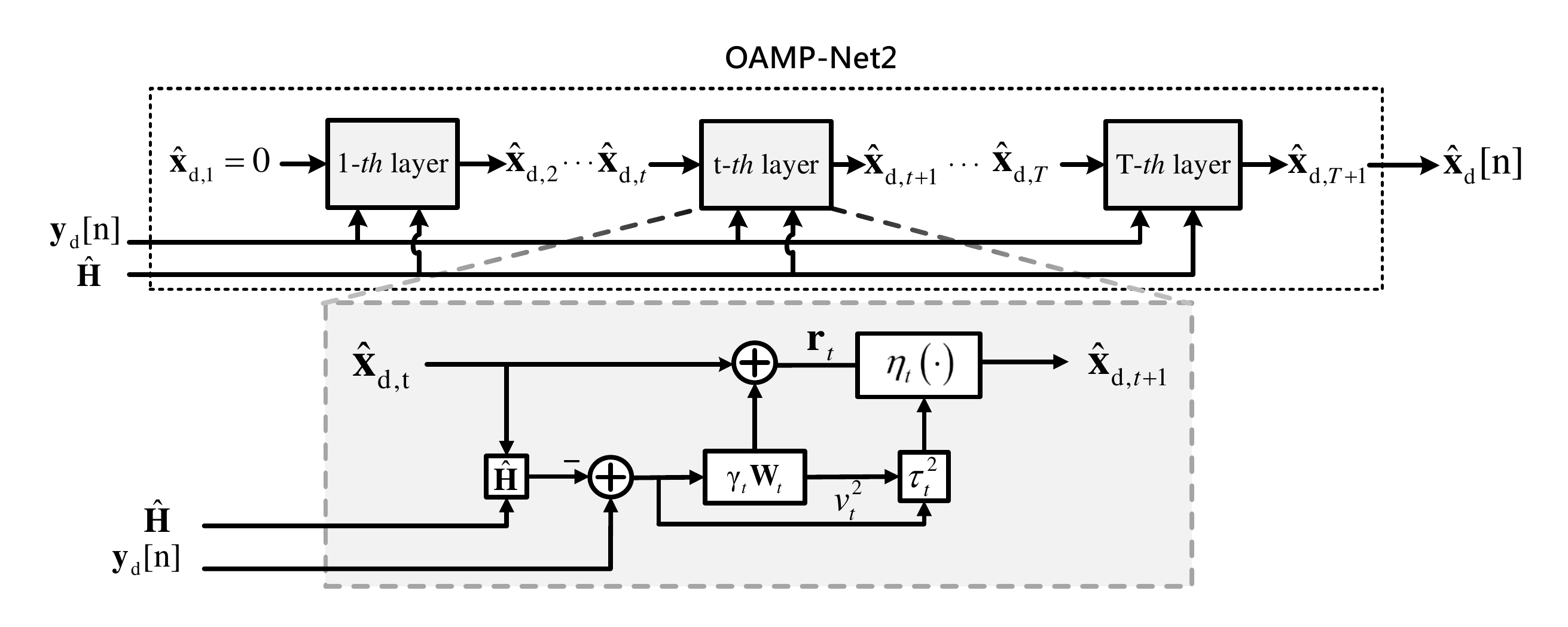}
  \caption{.~~Block diagram of OAMP-Net2 detector. The network consists of $T$ cascade layers, and each layer has the same structure that contains the linear estimator $\mathbf{W}_{t}$, nonlinear estimator $\eta_{t}(\cdot)$, error variance $\tau^{2}_{t}$ and  $v^{2}_{t}$, and tied weights.}\label{LOAMP_new}
\end{figure*}
\subsection{OAMP-Net2 Detector}
In this section, we introduce the OAMP-Net2 detector, which is developed by unfolding the OAMP detector. After presenting the network architecture, we elaborate the linear and nonlinear estimators, error estimators, and learnable variables in OAMP-Net2 detector.
\subsubsection{Network Architecture}\label{EP-Net}
The structure of  the OAMP-Net2 is illustrated in Fig.\,\ref{LOAMP_new}, which is obtained by unfolding the OAMP detector and adding several trainable parameters. The input of the OAMP-Net2 is the received signal vector, $\by_{\rmd}$, and the estimated channel matrix, $\hat{\bH}$,  while the final output is $\hat{\mathbf{x}}_{\rmd,T+1}$.

The network consists of $T$ cascade layers and each has the same architecture but different trainable parameters. For the $t$-th layer of the OAMP-Net2, the input is the estimated signal $\hat{\mathbf{x}}_{\rmd,t}$ from the ($t-1$)-th layer. Signal detection is performed as follows
\begin{equation}\label{eqlr}
  \mathbf{r}_{t}=\hat{\mathbf{x}}_{\rmd,t}+\gamma_{t}\mathbf{W}_{t}(\mathbf{y}_{\rmd}-\hat{\mathbf{H}}\hat{\mathbf{x}}_{\rmd,t}),
\end{equation}
\begin{equation}\label{eqldiver}
  \hat{\mathbf{x}}_{\rmd,t+1} = \eta_{t}(\mathbf{r}_{t},\tau^{2}_{t}; \phi_{t}, \xi_{t}),
\end{equation}
\begin{equation}\label{eqlv}
  v_{t}^{2}=\frac{\|\by-\hat{\bH}\hat{\bx}_{\rmd,t}\|_{2}^{2}-\mathrm{tr}(\bR_{\hat{\bn}_{\rmd}\hat{\bn}_{\rmd}})}{\mathrm{tr}(\hat{\bH}^{H}\hat{\bH})},
\end{equation}
\begin{equation}\label{eqlt}
  \tau^{2}_{t}=\frac{1}{\Nt}\mathrm{tr}(\mathbf{C}_{t}\mathbf{C}_{t}^{H})v_{t}^{2}+\frac{\theta_{t}^{2}}{\Nt}\mathrm{tr}(\mathbf{W}_{t}\bR_{\hat{\bn}_{\rmd}\hat{\bn}_{\rmd}}\mathbf{W}_{t}^{H}).
\end{equation}

The critical difference between the OAMP and OAMP-Net2 detectors is there are learnable variables $\Omega_{t}=\{\gamma_{t},\phi_{t},\xi_{t},\theta_{t}\}$ in each layer. By optimizing these parameters in the training process, the detection performance can be improved.
When $\gamma_{t} = \theta_{t} = 1$, $\phi_{t} = 1$ and $\xi_{t} = 0$, the OAMP-Net2 is reduced to the OAMP detector. The matrix $\bC_{t}=\bI-\theta_{t}\bW_{t}\hat{\bH}$ in the OAMP-Net2 is a revised structure of $\bB_{t}$ in the OAMP detector with learnable parameter $\theta_{t}$  to regulate the error variance $\tau_{t}^{2}$. We will introduce  the linear estimator, $\bW_{t}$, nonlinear estimator, $\eta_{t}(\cdot)$, error estimators, $\tau^{2}_{t}$ and  $v^{2}_{t}$, and learnable variables, $\Omega_{t}$, respectively.

\subsubsection{Linear and Nonlinear Estimator Modules}
The linear estimator module in (\ref{eqlr}) is a revised structure of (\ref{eqr}) in the OAMP algorithm by adding corresponding learnable parameter $\gamma_{t}$.
The learnable parameter $\gamma_{t}$ can be interpreted as the step-size for the update of $\mathbf{r}_{t}$,  and  is $\gamma_{t} =1 $ in each iteration in the OAMP detector. The de-correlated matrix $\bW_{t}$ is given in (\ref{UMMSE}).
The nonlinear estimator $\eta_{t}(\cdot)$ in OAMP-Net2 is constructed by the divergence-free estimator
\begin{equation}\label{divergence_free}
  \eta_{t}(\mathbf{r}_{t},\tau^{2}_{t}; \phi_{t},\xi_{t}) = \phi_{t}(\mathtt{E}\left\{\mathbf{x}_{\rmd}|\mathbf{r}_{t},\tau_{t}\right\}-\xi_{t}\mathbf{r}_{t}),
\end{equation}
where $\mathtt{E}\left\{\mathbf{x}|\mathbf{r}_{t},\tau_{t}\right\}$ is the MMSE estimate of $\mathbf{x}_{\rmd}$ with the equivalent AWGN channel (\ref{eqAWGN}), and has the same expression with (\ref{eqE}).  The nonlinear estimator, $\eta_{t}(\mathbf{r}_{t},\tau^{2}_{t}; \phi_{t}, \xi_{t})$, can be interpreted as
a linear combination of the  priori mean, $\mathbf{r}_{t}$, and the posteriori mean, $\mathtt{E}\left\{\mathbf{x}_{\mathrm{d}}|\mathbf{r}_{t},\tau_{t}\right\}$, which has been investigated in \cite{OAMP} and \cite{DOAMP}.

Compared with the MMSE estimator (\ref{eqE}) in the OAMP detector, the divergence-free estimator (\ref{divergence_free}) considers the contribution of the linear estimator and learnable parameters $(\phi_{t},\xi_{t})$. 
The MMSE estimator (\ref{eqE}) can be interpreted as a special case of  (\ref{divergence_free}) by setting $\phi_{t}=1$ and $\xi_{t}=0$.
\subsubsection{Error Variance Estimators}
The error variance estimators $v^{2}_{t}$ in (\ref{eqlv}) and $\tau^{2}_{t}$ in (\ref{eqlt})  play  important roles in providing appropriate variance estimates required for the linear and nonlinear estimators in the OAMP-Net2 detector. For error variance $v_{t}^{2}$, we adopt the same estimator with the OAMP detector in (\ref{eqlv}). For error variance $\tau_{t}^{2}$,
we construct the estimator based on  two assumptions on  error vectors $\bp_{t}$ and $\bq_{t}$ in \cite{OAMP} and incorporate learnable variables ($\gamma_{t}, \theta_{t}$). The detailed derivation for the two variance estimators $v^{2}_{t}$ and $\tau^{2}_{t}$ is provided in Appendix B. Furthermore, we substitute $v_{t}^{2}$ by $\mathrm{max}(v_{t}^{2},\epsilon)$  for a small positive constant $\varepsilon = 5\times10^{-13}$ to avoid stability problem.

\subsubsection{Learnable Variables}
The learnable variables $\mathbf{\Omega} = \{\Omega_{t}\}_{t=1}^{T}$ are optimized in the training process for OAMP-Net2. The Bayes-optimal property of the OAMP algorithm  has been proven in \cite{EP17Takeuchi}, but
it is derived in the large system with unitarily-invariant matrix $\bH$. In fact, the performance of the OAMP detector is far from the Bayes-optimal performance in practical finite-dimensional MIMO systems, especially when there are strong spatial correlation and channel estimation error. These observations motivate us to improve the original iterative detector with several trainable parameters  to adapt to various channel environments.

Similar to the TPG detector \cite{TPG18} and OAMP-Net\cite{OAMP_Net}, the OAMP-Net2 uses two learnable parameters $(\gamma_{t},\theta_{t})$  to adjust the linear estimator, and to provide appropriate stepsize for the update of the mean $\br_{t}$ and variance $\tau_{t}^{2}$ in the MMSE estimator.  On the other hand, the linear estimator (\ref{eqlr}) is related to gradient descent algorithm and its convergence behavior and performance are determined by appropriate step-size  of moving to the search point. The optimal step-size $\gamma_{t}$ can be learned  from the data for the update of the prior mean $\br_{t}$ in the MMSE estimator. Furthermore, the parameter $\theta_{t}$ has the similar function for the error variance $\tau^{2}_{t}$, which can compensate for the channel estimation error and regulate the $\tau^{2}_{t}$ to  provide appropriate value for the update of the prior variance in the MMSE estimator.

The parameters ($\phi_{t}$, $\xi_{t}$) in the nonlinear estimator $\eta_{t}(\cdot)$ play important roles in constructing an appropriate divergence-free estimator, which has been discussed in \cite{OAMP}. In precise, the divergence-free estimator (\ref{divergence_free}) can be applied in the OAMP detector, but the $\phi_{t}$ and $\xi_{t}$ are related to the prior distribution of the original signal and difficult to calculate. Therefore, MMSE estimator (\ref{eqE}) is considered for simplicity in the OAMP detector.
In our early work \cite{OAMP_Net}, we set $\phi_{t}=1$ and $\xi_{t}=0$ in OAMP-Net and use  MMSE estimator (\ref{eqE}). By contrary, we adaptively learn two parameters $\phi_{t}$ and $\xi_{t}$ in OAMP-Net2 to construct the nonlinear estimator $\eta_{t}(\cdot)$  satisfying the divergence-free property.

\subsubsection{Complexity Analysis}
\begin{table*}[t]	
\centering
	\caption{\\ COMPUTATIONAL COMPLEXITY OF DIFFERENT DETECTORS}
	\label{tab:mimo}
    \begin{tabular}{@{}lcccccc@{}}
    \toprule
    \diagbox{Complexity}{Detectors}& OAMP-Net2&DetNet&DNN-dBP&DNN-MPD&TPG&LcgNet\\
    \midrule
    Computational  complexity &$O(TN_{t}^3)$&$O(TN_{t}^2)$&$O(TN_{r}N_{t})$&$O(TN_{r}N_{t})$&$O(TN_{r}N_{t})$&$O(TN_{r}N_{t})$\\
    \hline
    Learnable  variables &$4T$&$(6N_{r}N_{t}+2N_{r}+N_{t})T$&$T$&$(N_{r}+N_{t})T$&$2T$&$4N_{t}T$\\
    \bottomrule
    \end{tabular}
\end{table*}\label{complexity_analysis}

The computational complexity required for the OAMP-Net2 is $O(TN_{t}^3)$. Similar to the OAMP detector and OAMP-Net, the computational complexity is dominant by the matrix inverse in each layer in $(\ref{BMMSE})$. When $N_{t}$ is  relatively small (e.g., $4$ or $8$), the matrix inverse operation is always acceptable. By contrary,  the AMP detector \cite{AMP} has a complexity of $O(TN_{t}^2)$,  which is dominated by the matrix multiplication, but has performance deterioration in small-size MIMO systems. The LMMSE detector has a complexity of $O(N_{t}^3)$ as one matrix inversion is needed, but it is non-iterative and no training is required. Although ML can achieve the optimal performance, it has a complexity of
$O(|\mathcal{S}|^{N_{t}})$. In Table\,\ref{tab:mimo}, we compare the computational complexity of the OAMP-net2 detector with the state-of-art DL-based MIMO detectors  in \cite{DeepMIMO,Improving,CHemp18Tan,TPG18,GDNet} including DetNet \cite{DeepMIMO}, DNN-dBP \cite{Improving}, DNN-MPD \cite{CHemp18Tan}, TPG \cite{TPG18}, LcgNet \cite{GDNet}.
Although  the state-of-art DL-based MIMO detectors  have lower complexity than the OAMP-Net2 detector, their performance is deteriorated in the small-MIMO systems to some extent.

Furthermore, we investigate the number of learnable variables in different DL-based MIMO detectors. From  Fig.\,2, the total number of trainable variables is equal to $4T$ since each layer of the OAMP-Net2 contains
only four trainable variables $\Omega_{t} = (\gamma_{t},\phi_{t},\xi_{t},\theta_{t})$. By contrast, $2T$ trainable variables $(\gamma_{t},\theta_{t})$ are required to train in OAMP-Net. However, the numbers of learnable variables in DetNet, DNN-MPD and LcgNet heavily are dominated by the number of antennas in the transmitter or the receiver. By contrary, the number of trainable variables of the OAMP-Net2 and OAMP-Net are independent of the number of antennas $N_{r}$ and $N_{t}$, and only determined by the number of layers $T$. This is an advantageous feature for large-scale problems, such as massive MIMO detection. With only few trainable variables, the stability and speed of convergence can be improved in the training process.

\subsection{Soft-Input and Soft-Output}
 As many modern digital communication systems need to produce a probabilistic estimation of the transmitted data
given the observations to probabilistic channel decoder. A significant issue is whether the MIMO detector can use the soft-information from the decoder and produce the soft-output. Different from the DetNet in \cite{DeepMIMO} that only can  provide the soft-output, our proposed  detector are the soft-input and soft-output receiver and therefore can achieve the turbo equalization. We only provide the principle of the OAMP-Net2-based turbo receiver and the specific experimental results are outside the scope of this paper and will be conducted in the future.

From  ($\ref{divergence_free}$), the OAMP-Net2 can decouple the joint posterior probability $\mathcal{P}(\mathbf{x}|\mathbf{y}_{\rmd},\hat{\bH})$ into a series of marginal posterior probability $\mathcal{P}(x_{j}|\mathbf{r}_{t},\tau_{t})$. The marginal posterior probability is used to
produce the soft output  log-likelihood ratios (LLR), which is given by
\begin{equation}\label{LLR}
L_{A}(b_{j,k})=\log\frac{\sum_{\mathcal{S}_{j}^{+}}\mathcal{P}(x_{j}|\mathbf{r}_{t},\tau_{t})}{\sum_{\mathcal{S}_{j}^{-}}\mathcal{P}(x_{j}|\mathbf{r}_{t},\tau_{t})},
\end{equation}
where $b_{j,k}$ is the $k$-th bit in the transmitted symbol $x_{j}$, and $\mathcal{S}_{j}^{+}$ and $\mathcal{S}_{j}^{-}$ denote the subsets of the constellation symbols with the $k$-th bit being $1$ and $0$, respectively. After interleaved and delivered to the channel encoder, extrinsic LLR can be computed and given to the OAMP-Net2 detector as updated prior information. The detector and decoder iteratively exchange information until convergence.
\subsection{Practical Implementation}
The developed  detector can be divided into two stages. In the offline training stage, we obtain the optimized parameters,  $\mathbf{\Omega}=\{\Omega_{t}\}_{t=1}^{T}$,  for different SNRs and channel correlation coefficients based on TensorFlow platform and GPUs' powerful computing ability. The optimized parameters will be stored to detect the modulated symbols in the deployment stage. The OAMP-Net2 detector can be interpreted as a new iterative detector after training. The incorporated learnable parameters can adapt to practical channel and compensate for channel estimation error to improve the detection performance. Although aforementioned implementation process is in an offline manner, the proposed detector can also be implemented by online training to adapt to the fluctuations in the channel conditions owing to the superiority of its low demand for training data and computational resources. 
\section{Numerical Results}\label{Simulation}
In this section, we provide  numerical results to show the  performance of the proposed model-driven detector for MIMO detection. First, we elaborate the implementation details and parameter settings. Then, the performance of the OAMP-Net2 is presented under i.i.d. and correlated Rayleigh  MIMO channels with perfect CSI. Afterwards,  we investigate the performance of the OAMP-Net2-based JCESD architecture. Finally, the robustness of the OAMP-Net2 to SNR and channel correlation is demonstrated.
\subsection{Implementation Details}
In our simulation, OAMP-Net2 is implemented in Tensorflow by using a PC with GPU NVIDIA GeForce GTX 1080 Ti. The SNR of the system, defined as
\begin{equation}\label{eqsnr}
  \mathrm{SNR}=\frac{\mathtt{E}\|\mathbf{H}\mathbf{x}_{\rmd}\|^{2}_{2}}{\mathtt{E}\|\mathbf{n}_{\rmd}\|^{2}_{2}},
\end{equation}
is used to measure the noise level. We assume the same SNR in pilot and data transmission stage. Each element of the i.i.d. Rayleigh MIMO channel $\bH$ is  $h_{i,j} \sim \mathcal{N}_{\mathbb{C}}(0,1/\Nr)$ while the correlated Rayleigh MIMO channel is described by the Kronecker model,
\begin{equation}\label{eqcor}
  \mathbf{H}_{c}=\mathbf{R}_R^{1/2}\mathbf{G}\mathbf{R}_T^{1/2},
\end{equation}
where $\mathbf{R}_{R}$ and $\mathbf{R}_{T}$ are the spatial correlation matrix at the receiver and the transmitter, respectively, and are generated according to the exponential correlation model \cite{corMIMO} with the same correlation coefficient $\rho$, and $\mathbf{G}$ is the i.i.d. Rayleigh fading channel matrix. Furthermore, to implement OAMP-Net2 in the Tensorflow, we consider equivalent real-valued representation for all aforementioned problem models.

The training data consists of a number of randomly generated pairs $\bd^{(i)}\triangleq(\bx^{(i)},\by^{(i)})$. For each pair $\bd^{(i)}$, channel $\bH$ is randomly generated from the i.i.d. or correlated Rayleigh MIMO channel model. The data $\mathbf{x}^{(i)}$ is generated from the $M$-QAM modulation symbol with $M$ being the  modulation order. We train the network with $1,000$ epochs. At each epoch, the training set contains $5,000$  different samples $\mathbf{d}^{(i)}$, and $1,000$  different validation samples. The validation sets are used to choose the best network for each epoch in the training stage. In the test stage, we generate the  data to test the network until the number of bit errors exceed $10,000$. The network is trained using the stochastic gradient descent method and Adam optimizer. The learning rate is set to be $0.001$ and the batch size is set to $100$. To train the learnable variables well, the learning rate is set to be $0.0001$ and validation samples is $10,000$ in high SNR regime (SNR = $30$ or above). Except for special instructions, we choose the $l_{2}$ loss \footnote{We train the OAMP-Net2 with $l_{2}$ loss and cross-entropy, respectively, and find the network trained with $l_{2}$ loss outperforms that trained with cross-entropy.} as the cost function in all experiment settings, which is defined as
\begin{equation}\label{wploss}
l_{2}(\mathbf{\Omega})=\frac{1}{D}\sum_{\mathbf{d}^{(i)}\in \mathcal{D}}\|\mathbf{x}_{\rmd}^{(i)}-\hat{\mathbf{x}}_{\rmd,T+1}(\mathbf{y}^{(i)})\|^{2}_{2},
\end{equation}
where $\mathcal{D}$ is the training set of a mini-batch and $D = |\mathcal{D}|$.

\begin{figure}[!t]\centering	
\subfloat[QPSK]{\includegraphics[width=3in]{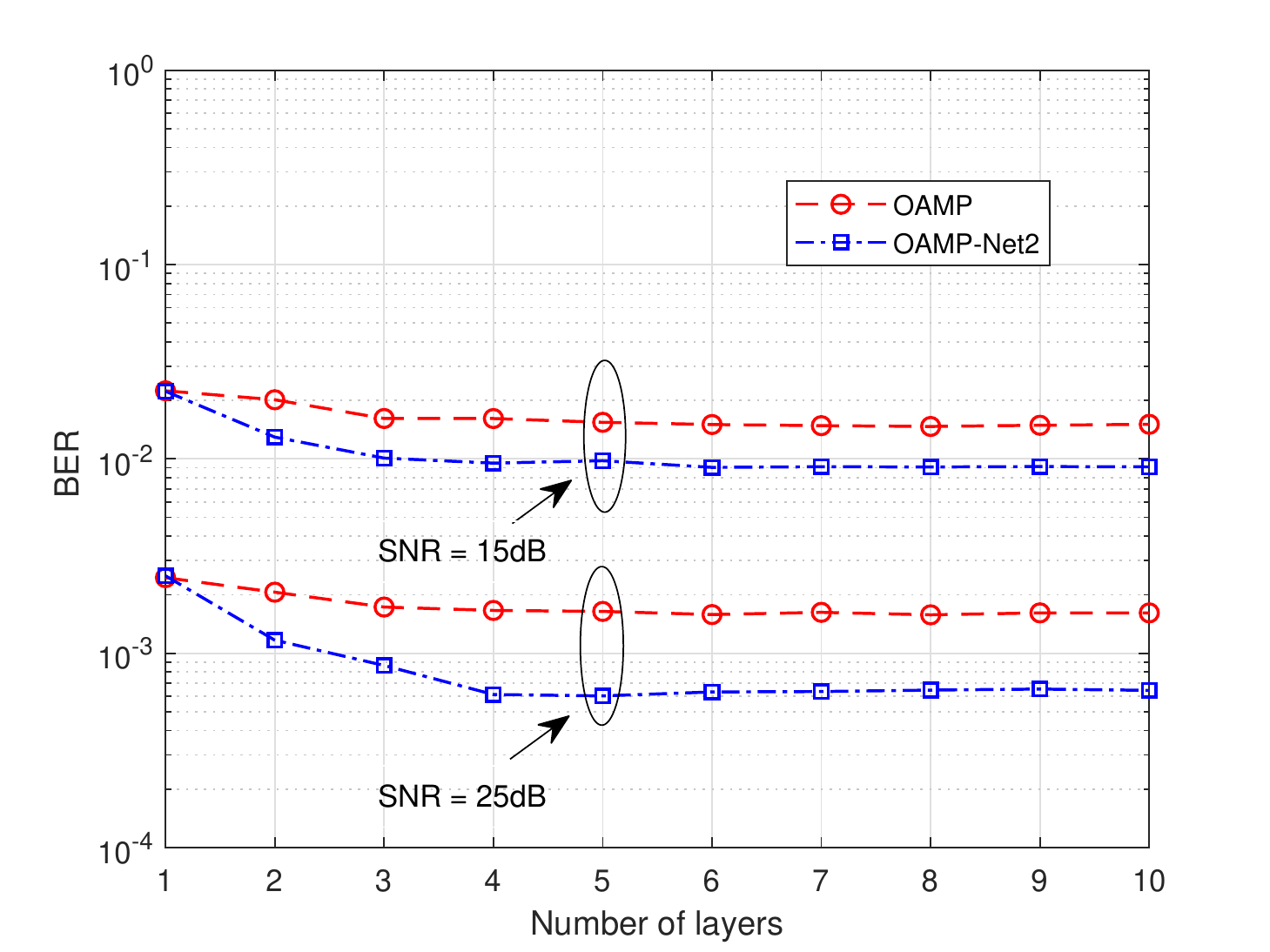}\label{fig5a}}\\
	\subfloat[$16$-QAM]{\includegraphics[width=3in]{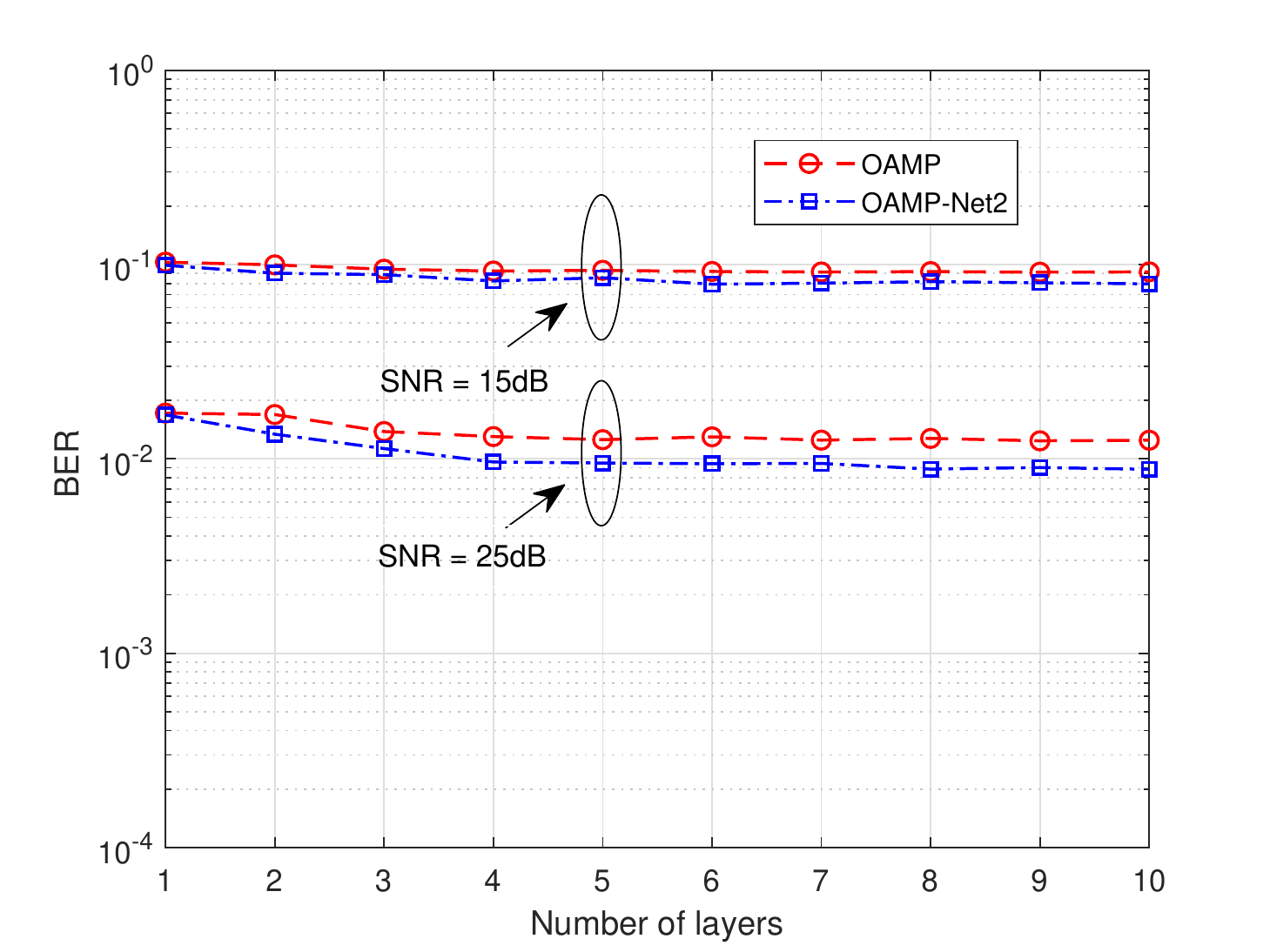}\label{fig5b}}
	\caption{.~~BERs performance of the OAMP-Net2 and OAMP detector versus the number of layers under QPSK and $16$-QAM modulation.}\label{convergence}
\end{figure}

In general, the optimal variables $\mathbf{\Omega}$ may be different for different SNRs. Therefore, we train the OAMP-Net2 for each SNR separately, which means that the training and test data are generated with the same fixed discrete SNR value. For the correlated Rayleigh MIMO channel, we train the network for each correlation coefficient $\rho$,  respectively. The network needs to be retrained if different modulation symbols are utilized. Furthermore, we also consider the robustness of the OAMP-Net2 to various mismatches.

\subsection{Perfect CSI}
In this section, we study the performance of OAMP-Net2 for MIMO detection with perfect CSI. After analyzing its convergence, the detection performance of the OAMP-Net2 under i.i.d. and  correlated Rayleigh MIMO channels is investigated.
\subsubsection{Convergence analysis}
First, we analyze the convergence  of the OAMP-Net2 and OAMP detectors. Fig.\,\ref{convergence} illustrates the bit-error rate (BER) performance versus the number of layers (iterations) under $\mathrm{SNR} = 15$dB and $25$dB with QPSK and $16$-QAM constellations. From Fig.\,\ref{convergence}, both the OAMP-Net2 and OAMP detectors converge within four layers (iterations) for all the cases and OAMP-Net2 can obtain more performance gain when SNR = $25$dB, which means the OAMP-Net2 detector improves OAMP detector significantly in high SNR regime. Furthermore, compared with $16$-QAM,  the performance gap between OAMP and OAMP-Net2  is larger for QPSK. Based on the above observations, we set OAMP and OAMP-Net2 detectors  four layers (T=4) in following simulation when SNR$\leqslant 25$ dB, and ten layers (T=10) when SNR = $30$dB or above except for specific instructions.

\begin{figure}[!t]\centering	
\subfloat[$4\times4$ MIMO]{\includegraphics[width=3in]{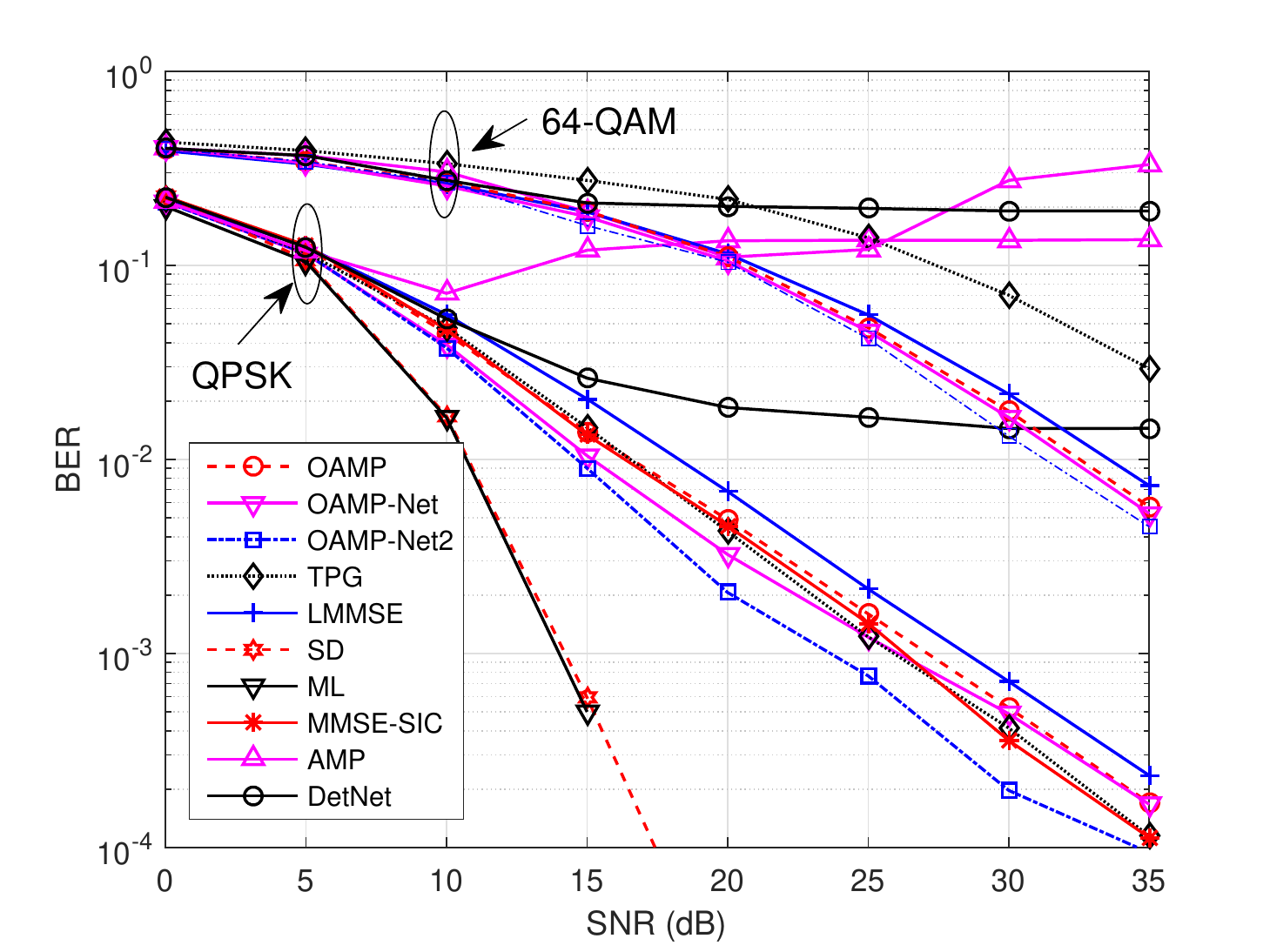}\label{fig5a}}\\
	\subfloat[$8\times8$ MIMO]{\includegraphics[width=3in]{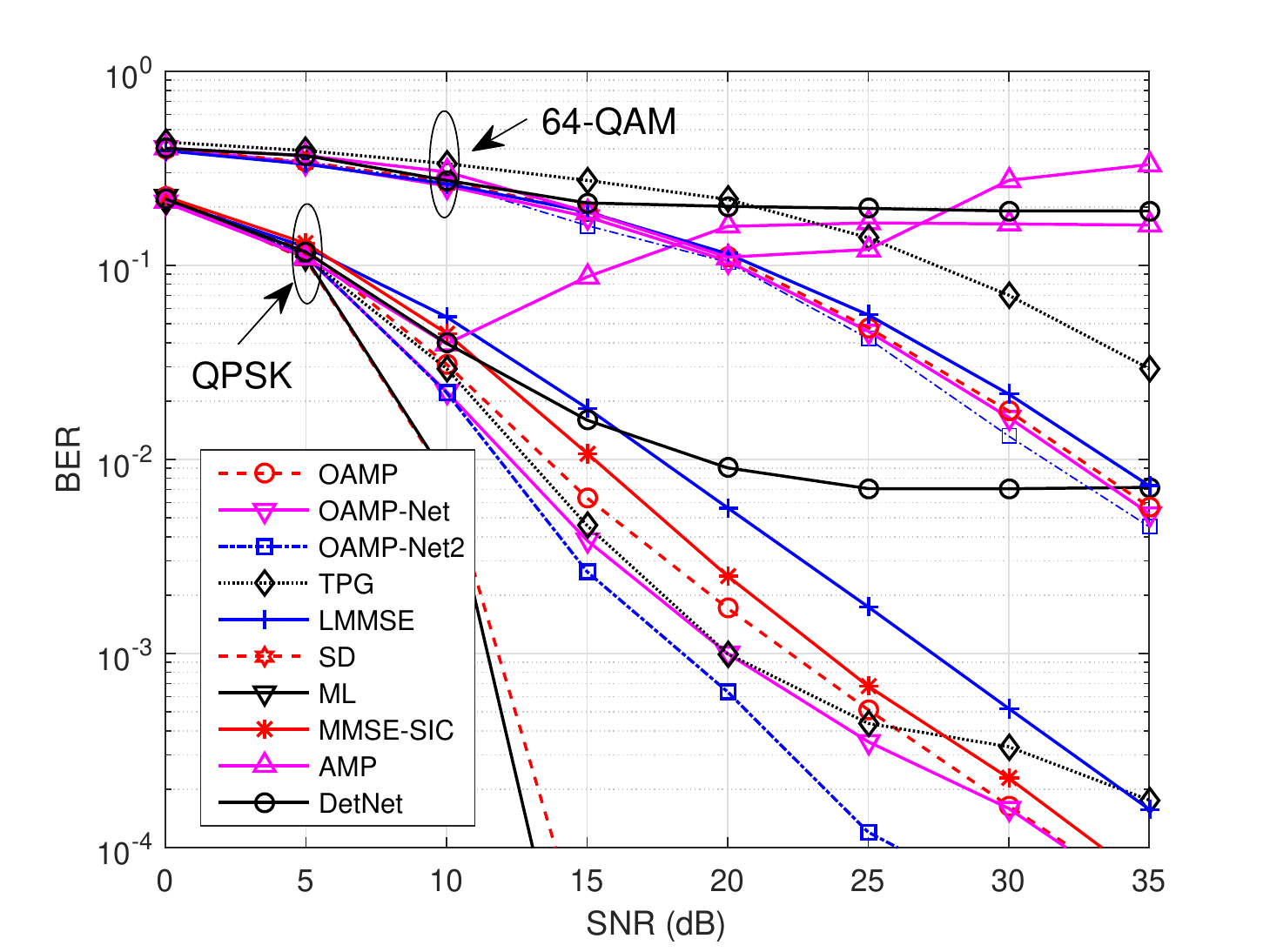}\label{fig5b}}
	\caption{.~~BERs performance comparison of the OAMP-Net2 with other MIMO detectors under i.i.d. Rayleigh MIMO channels with different orders of modulation.}\label{Rayleigh}
\end{figure}

\subsubsection{Performance Comparison}

\begin{figure}
\begin{minipage}{3in}
  \centerline{\includegraphics[width=3in]{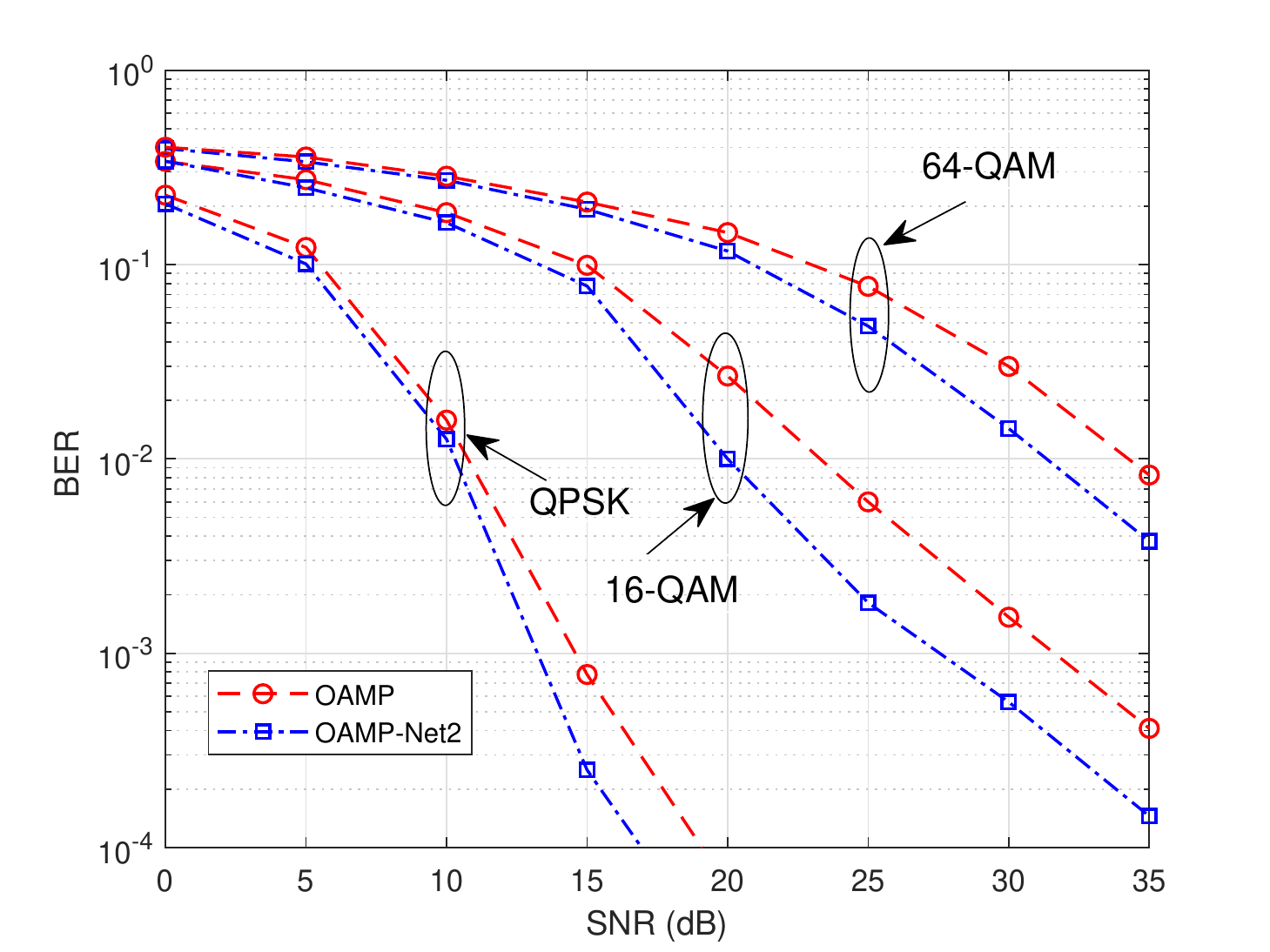}}
  \centerline{(a) $16\times16$ MIMO}
\end{minipage}
\hfill
\begin{minipage}{3in}
  \centerline{\includegraphics[width=3in]{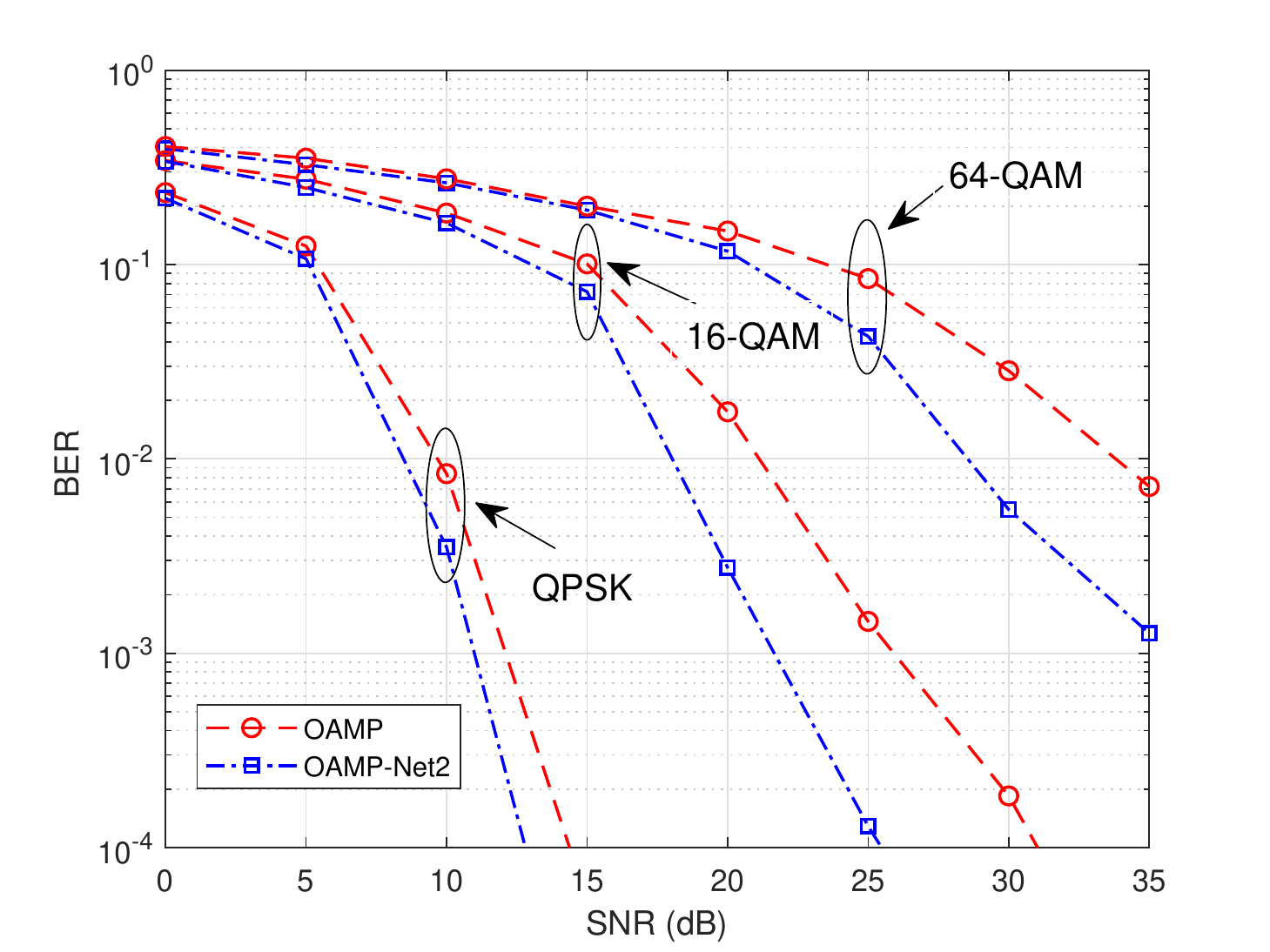}}
  \centerline{(b) $32\times32$ MIMO}
\end{minipage}
\caption{.~~BERs performance comparisons of the OAMP  and OAMP-Net2 under high-dimensional i.i.d. Rayleigh MIMO channels with different modulation symbols.}
\label{High_dimensional}
\end{figure}

 Fig.\,\ref{Rayleigh} compares the BER performance of the OAMP, OAMP-Net \cite{OAMP_Net}, TPG \cite{TPG18}, DetNet\cite{DeepMIMO}, AMP\cite{AMP}, SD, ML, MMSE-SIC \cite{MMSE-SIC}, LMMSE and OAMP-Net2 detectors. Both $4\times4$ and $8\times8$ MIMO systems are investigated. Among these detectors,  the TPG detector \cite{TPG18} is developed with  two trainable parameters in each layer and has a similar architecture to OAMP-Net. Furthermore, the OAMP and MMSE-SIC detectors can achieve soft decision. From the figure, the OAMP-Net2 outperforms other MIMO detectors except for SD and ML, but the computational complexity of the ML and SD detectors is prohibitive. The DetNet and AMP detectors have serious performance deterioration in small-size MIMO systems. Furthermore, the OAMP-Net, TPG, and OAMP-Net2 detectors outperform the OAMP detector in all setting, which demonstrates DL can improve the OAMP detector by learning the optimal parameters. But the performance gain is limited for $64$-QAM. Compared with OAMP-Net in \cite{OAMP_Net}, OAMP-Net2 has a significant performance improvement in all setting. The reason is the OAMP-Net2 can learn the optimal parameter $(\phi_{t},\xi_{t})$ to construct the nonlinear estimator satisfying the divergence-free property.
For high-dimensional MIMO systems in Fig.\,\ref{High_dimensional}, similar conclusions can be obtained.

\begin{figure}
\begin{minipage}{3in}
  \centerline{\includegraphics[width=3in]{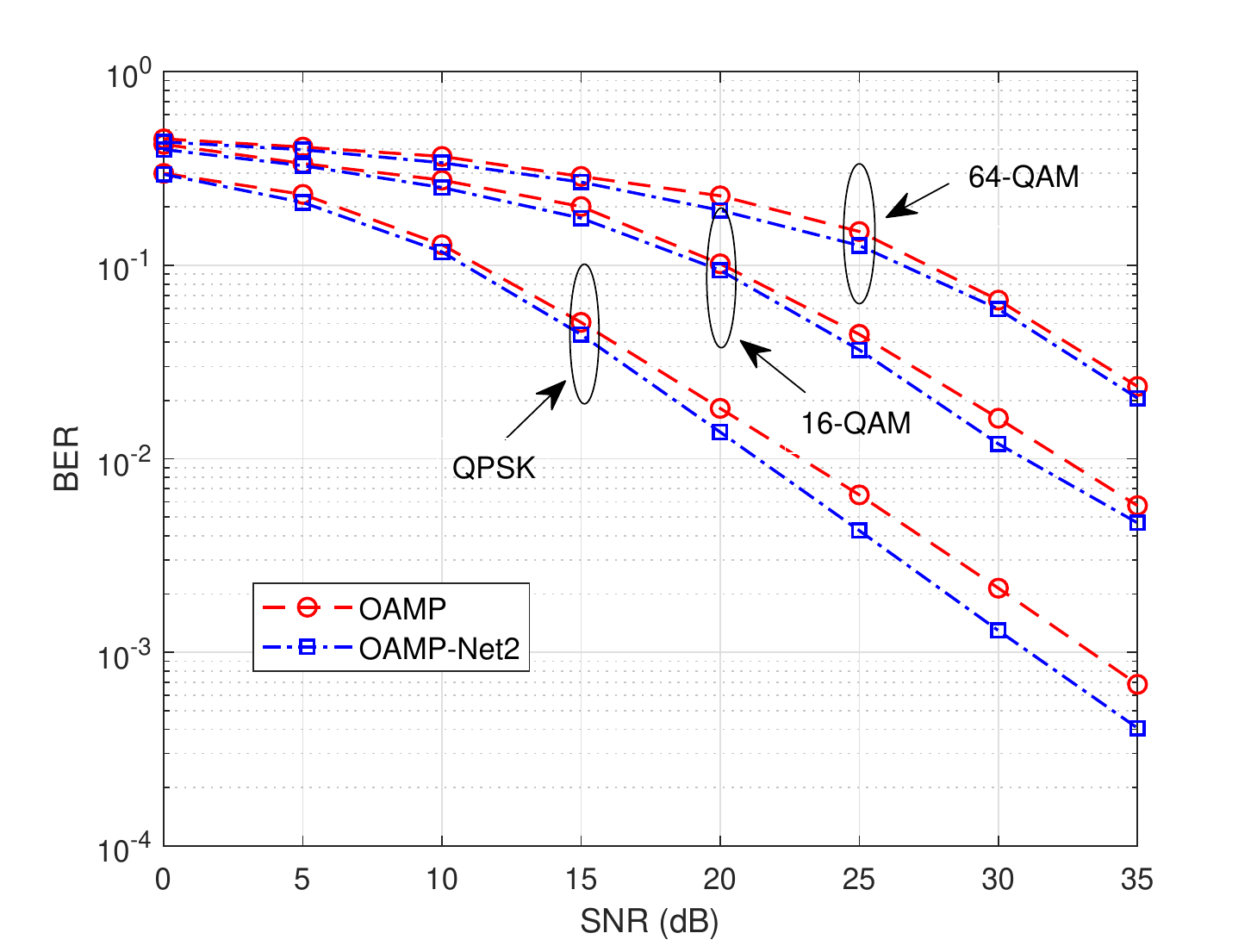}}
  \centerline{(a) $4\times4$ MIMO}
\end{minipage}
\hfill
\begin{minipage}{3in}
  \centerline{\includegraphics[width=3in]{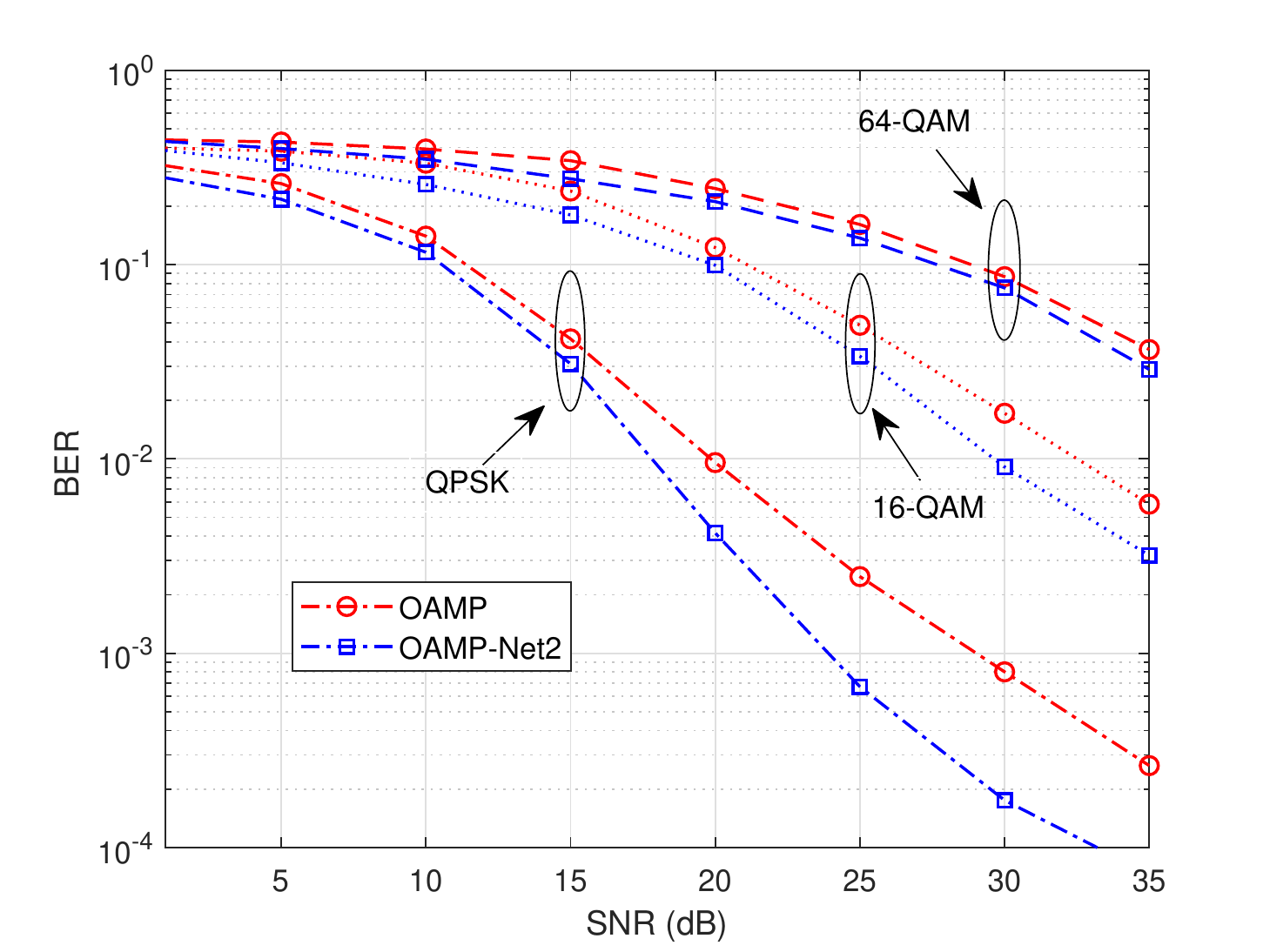}}
  \centerline{(b) $8\times8$ MIMO}
\end{minipage}
\caption{.~~BERs performance comparison of the OAMP  and OAMP-Net2 under correlated Rayleigh MIMO channels with $\rho=0.5 $ for different orders of modulation.}
\label{correlated}
\end{figure}

\subsubsection{Correlated MIMO Channel Performance}
 Fig.\,\ref{correlated} illustrates the BER  of the OAMP-Net2 and the OAMP detector under correlated Rayleigh MIMO channel with $\rho=0.5$. We set the OAMP and OAMP-Net2 detectors ten layers with  different orders of modulation. From the figure, all algorithms have performance degradation compared with the i.i.d. Rayleigh MIMO channel, 
But the OAMP-Net2 detector still have the performance improvement over the OAMP detector in all setting and can obtain more performance improvement in $ 8\times8 $ MIMO system. Specifically,  if we target for BER=$10^{-2}$  with QPSK modulation, BER performance improves about $1.8$ dB for $4\times4$ MIMO. By contrast, it is approximately $2.2$ dB for $8\times8$ MIMO. But the performance gain of the OAMP-Net2 detector decreases with the increase of the modulation order. For example, if we target  for BER=$10^{-2}$ in $4\times4$ MIMO, SNR improves about $1.5$ dB for QPSK modulation. By contrast, it decreases to approximately $1.1$ dB for $16$-QAM .
\subsubsection{3GPP MIMO Channels Performance}
\begin{figure}[h]
  \centering
  \includegraphics[width=3in]{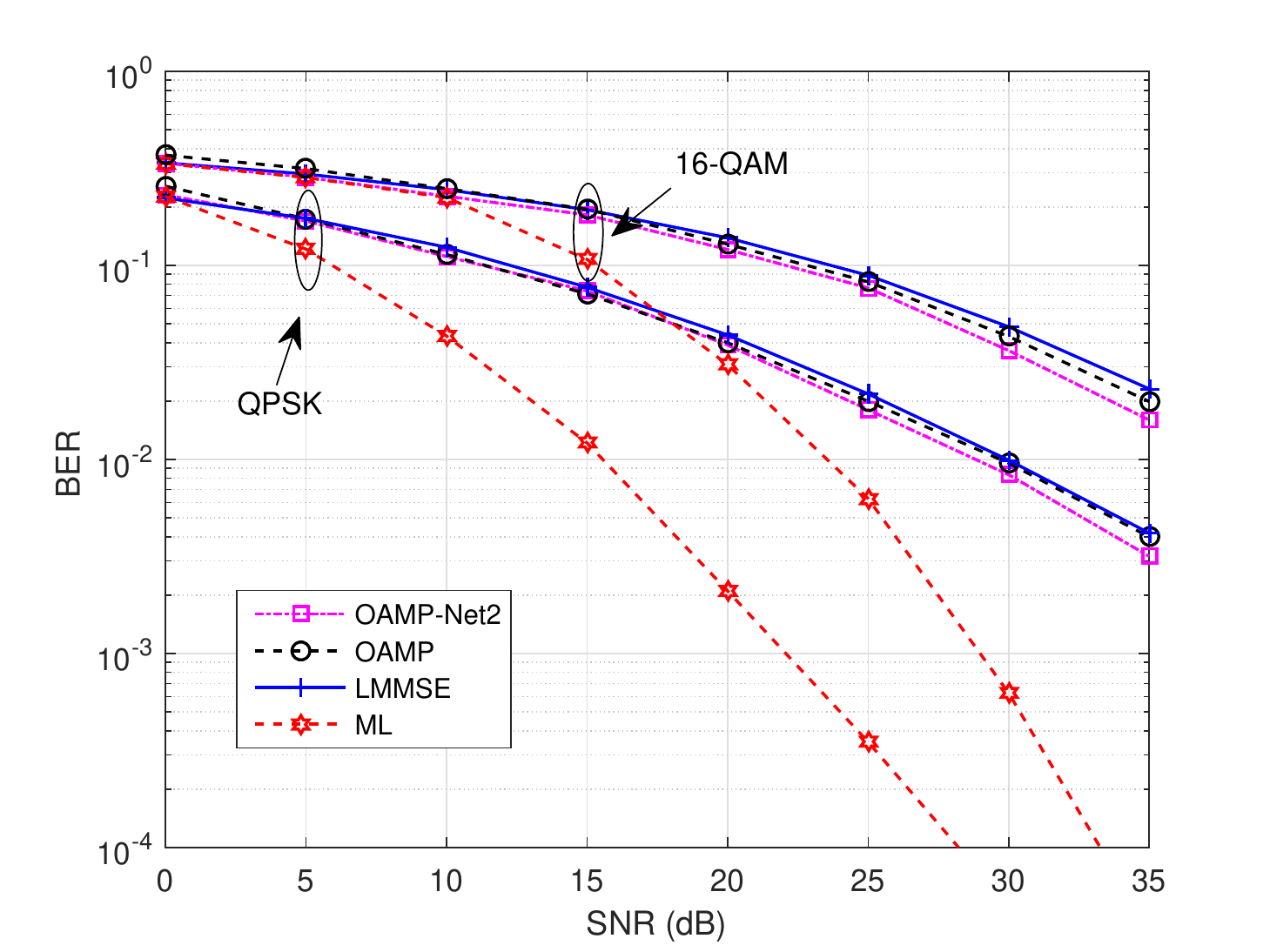}
  \caption{.~~BERs performance comparison of the OAMP-Net2 with other MIMO detectors under 3GPP MIMO channels with different orders of modulation.}\label{3GPP}
\end{figure}

In the aforementioned sections, we have investigated the OAMP-Net2 detector with Rayleigh fading channels. Here, we conduct numerical experiments by utilizing a dataset of the realistic channel samples from the 3GPP 3D MIMO channel \cite{3GPP}, as implemented in the QuaDRiGa channel simulator \cite{QuaDRiGa}. The channel parameters are set to be same as \cite{MMNet}. As shown in Fig.\,\ref{3GPP}, OAMP-Net2 indeed has performance loss with the realistic channel. The reason is that the OAMP-Net2 requires the channel matrix to be unitarily-invariant while the practical channel is very ill-conditioned and the channel correlation is serious. which incurs performance degradation. Furthermore, only $4$ trainable parameters in each layer may restrict the ability to adapt to realistic 3GPP MIMO channels. Nevertheless, the OAMP-Net2 still outperforms the OAMP and LMMSE detectors with limited performance gain.

\subsection{JCESD Performance}
 In the above, all detectors are investigated with accurate CSI. In this section,  we consider OAMP-Net2-based JCESD architecture for a $4\times4$ MIMO system. Both OAMP and OAMP-Net2 has four layers. In order to avoid gradient vanishing,  we use the summation of $l_{2}$-loss over all $L \times T$ layers to train the model, which is defined as
\begin{equation}\label{wploss}
l_{2}(\mathbf{\Omega})=\sum_{l=1}^{L}\sum_{t=1}^{T}\frac{1}{D}\sum_{\mathbf{d}^{(i)}\in \mathcal{D}}\|\mathbf{x}_{\rmd}^{(i)}-\hat{\mathbf{x}}_{\rmd,T}^{(l)}(\mathbf{y}^{(i)})\|^{2}_{2},
\end{equation}
and the learning rate is set to be $0.0001$.

\begin{figure}
  \centering
  \includegraphics[width=3in]{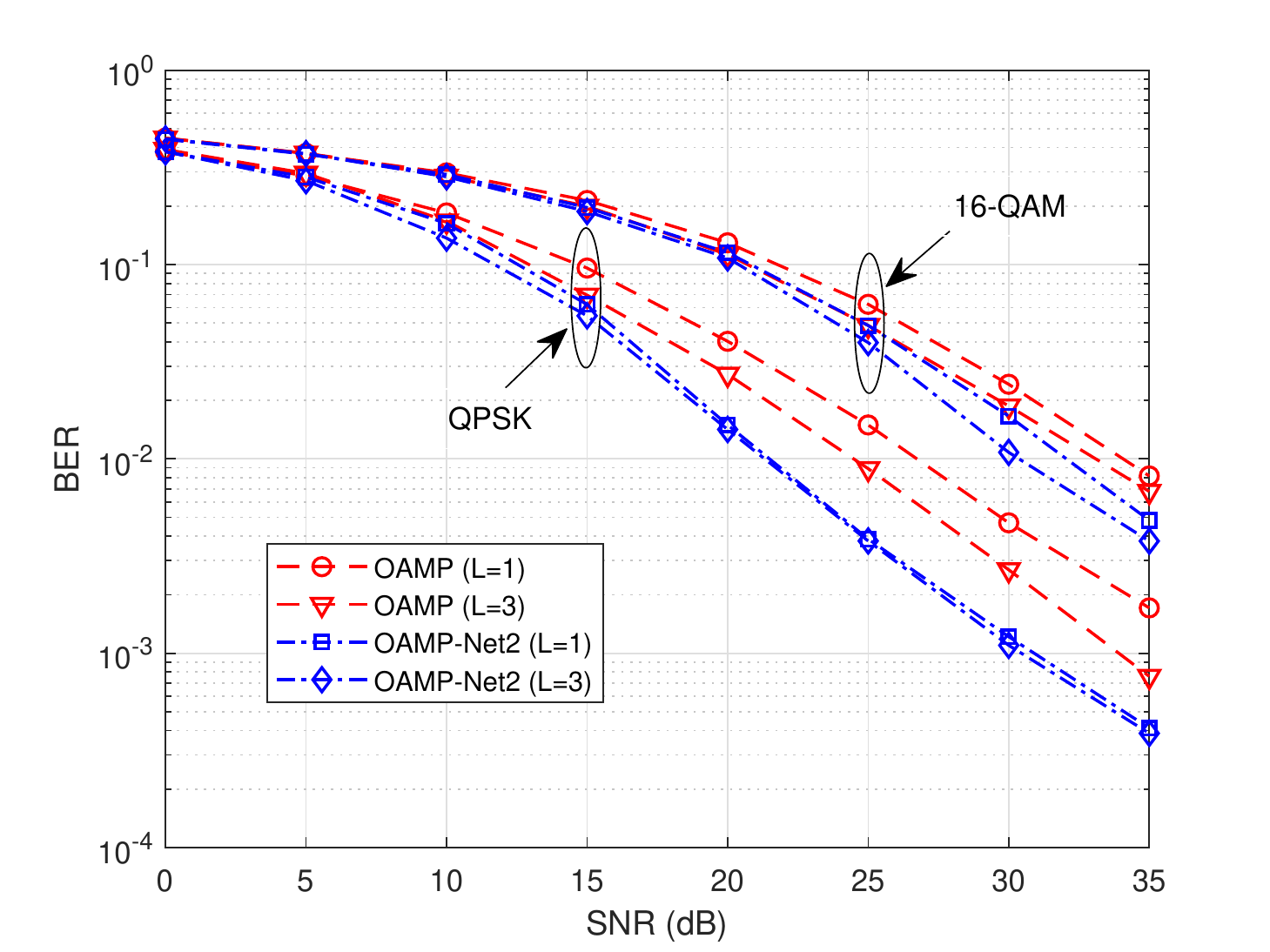}
  \caption{.~~BERs performance comparisons of the OAMP and OAMP-Net2 in JCESD architecture with $\Np=4$ and $\Nc=16$.}\label{JCESD_DFT}
\end{figure}
We consider orthogonal pilot in channel training stage, where the pilot matrix $\mathbf{X}_{\rmp} \in \mathbb{C}^{ \Nt\times \Np}$ is chosen by selecting $\Nt$ columns of the discrete Fourier transformation (DFT) matrix $\bF \in \mathbb{C}^{\Np\times \Np}$. Fig. \ref{JCESD_DFT} shows the BERs of the OAMP and OAMP-Net2 detectors in the JCESD architecture, where $L=1$ means no data feedback to channel estimator and $L=3$ means the detected data are feedback twice to channel estimator. From the figure, the OAMP-Net2 detector outperforms the OAMP detector in all settings significantly. Specifically, if we target  for BER=$10^{-2}$ with $16$-QAM, BER performance improves about $2.1$ dB for $L=1$ and improves about $2.9$ dB
for $L=3$. Furthermore, we observe that OAMP-Net2 detector without data feedback ($L=1$) outperforms OAMP detector with data feedback twice ($L=3$), which demonstrates learning some trainable variables can compensate for the channel estimation error to improve the equivalent SNR in detection stage. For QPSK modulation, the performance gain are $5.1$ dB and $3.1$ dB for $L=1$ and $L=3$, respectively, when we target  for BER=$10^{-2}$. In addition,  marginal performance is obtained by data feedback for the OAMP-Net2 detector with QPSK because learnable parameters have strong ability to compensate for channel estimation error. Therefore, no data feedback is needed in this case.
\subsection{Robustness}
In this section, we analyze the robustness of the OAMP-Net2 against various mismatches, including the SNR, channel correlation, MIMO configuration and modulation mismatches. As aforementioned numerical results are performed when training and test data are generated with same system parameters, an interesting question is whether the trained parameters are robust to various mismatches. Because of limited data and computing resources are available for online training, verifying the robustness of offline-trained network is particularly meaningful. 
\begin{figure}
  \centering
  \includegraphics[width=3in]{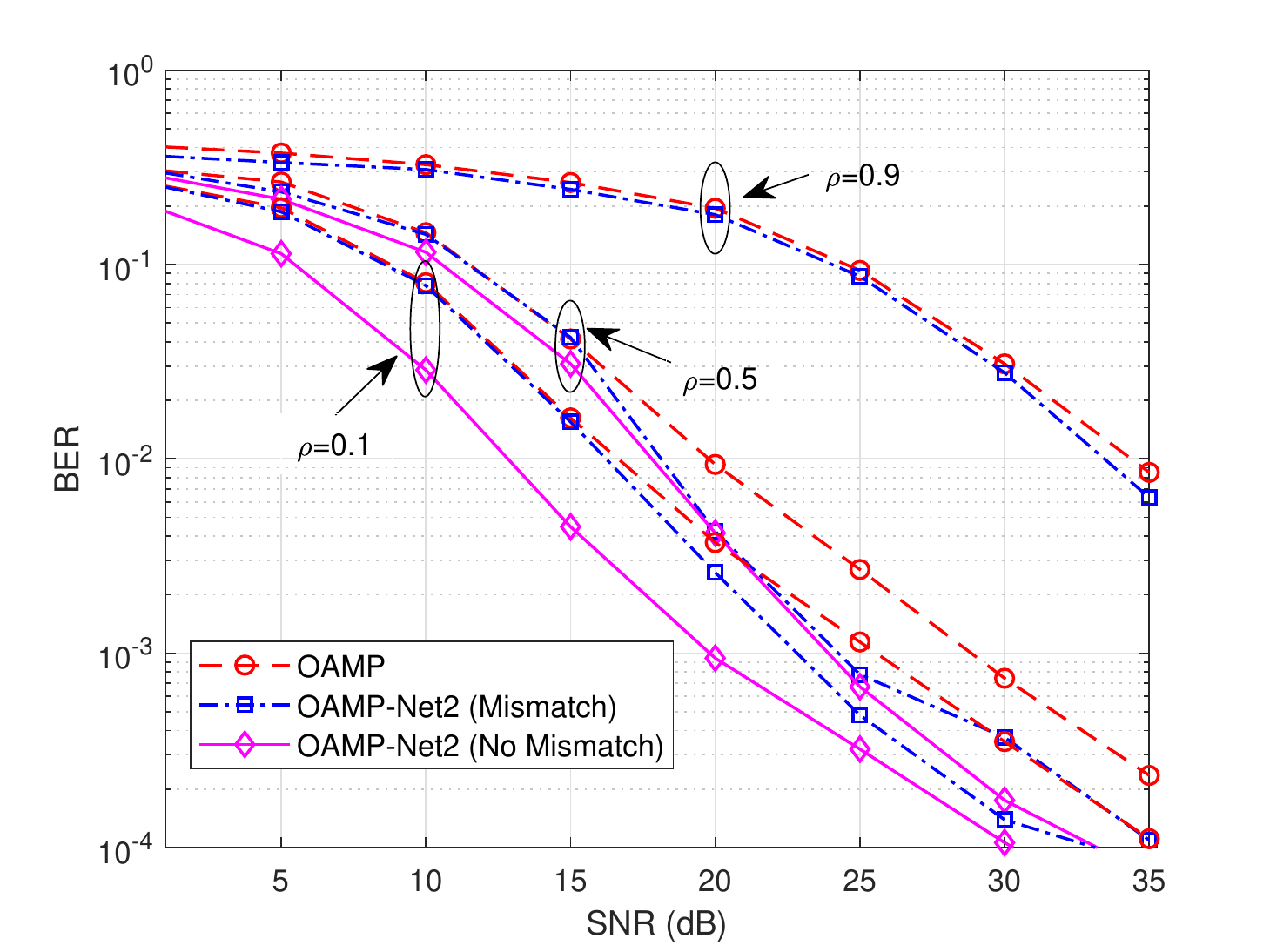}
  \caption{.~~BERs performance comparisons of the OAMP and OAMP-Net2 with SNR and correlation mismatches for QPSK modulation.}\label{mismatch1}
\end{figure}

\subsubsection{Robustness to SNR and Channel Correlation}
Fig.\,\ref{mismatch1} presents the BER performance of  OAMP and OAMP-Net2 with SNR and correlation mismatches in $8\times8$ MIMO system.
 The network is trained in the correlated Rayleigh MIMO channel (\ref{eqcor}) with channel correlation coefficient $\rho=0.5$ and SNR=$20$dB, and tested with mismatched channel correlation coefficient $\rho$ and SNR, which are shown in the figure. From the figure, the trained network with correlation mismatch still outperforms the OAMP detector significantly in all setting and has $0.3$ dB performance loss, if we target for BER=$10^{-2}$ and $\rho=0.5$, which demonstrates the OAMP-Net2 has strong robustness to mismatch. Interestingly, the trained network even outperforms the OAMP detector with $\rho =0.1 $ in high SNR regime (SNR=$20$-$30$ dB). Since the learnable variables successfully compensates for the disadvantages of channel correlation. Compared with perfect SNR\footnote{Perfect SNR  means that network in the training and test stage have the same SNR.}, the OAMP-Net2 detector with SNR mismatch has little performance deterioration except for SNR = $25$dB or above, when we target $\rho=0.5$.

\subsubsection{Robustness to MIMO Configuration and Modulation}

\begin{figure}
\begin{minipage}{3in}
  \centerline{\includegraphics[width=3in]{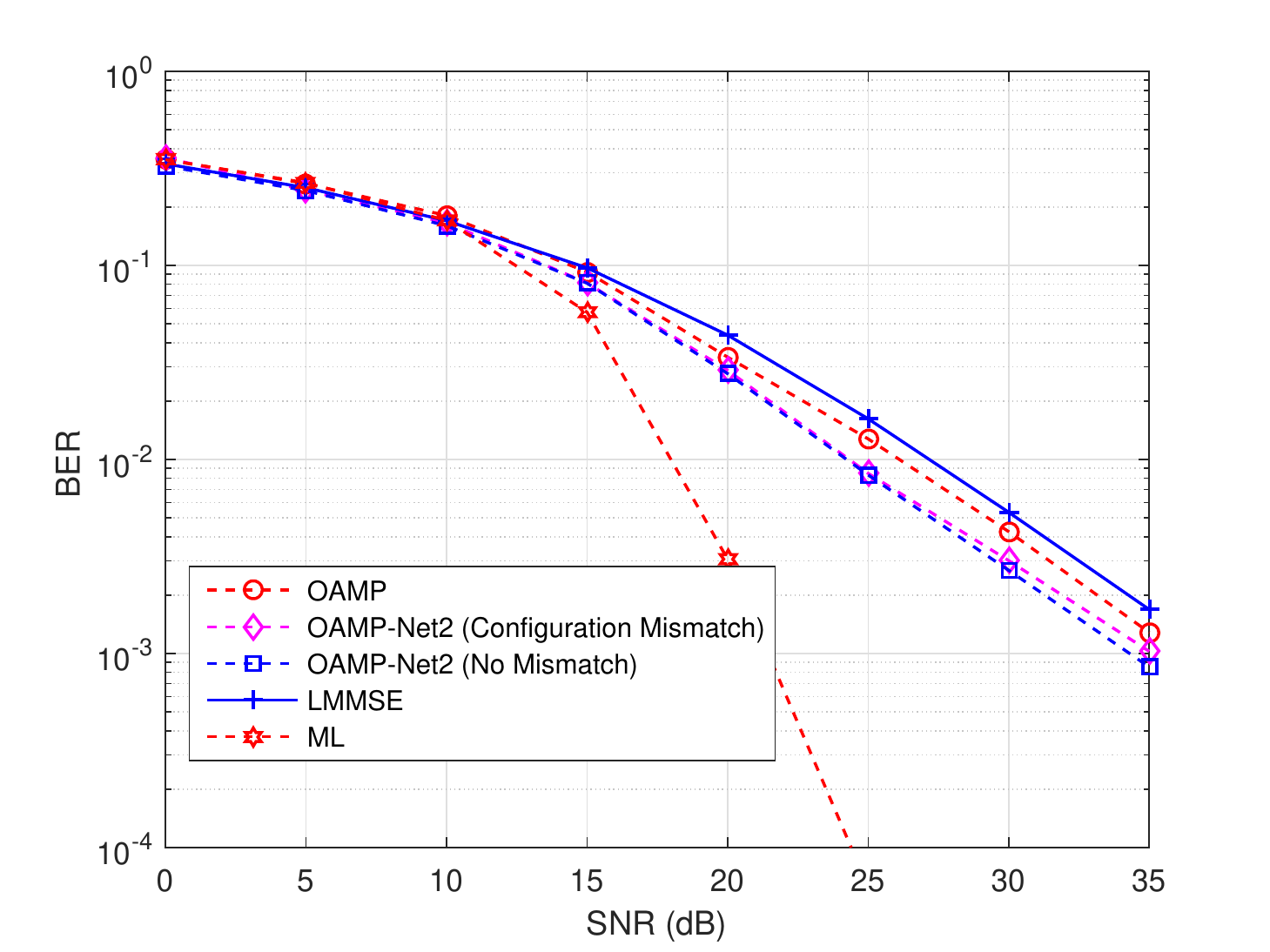}}
  \centerline{(a) MIMO configuration mismatch}
\end{minipage}\label{Robustness2a}
\hfill
\begin{minipage}{3in}
  \centerline{\includegraphics[width=3in]{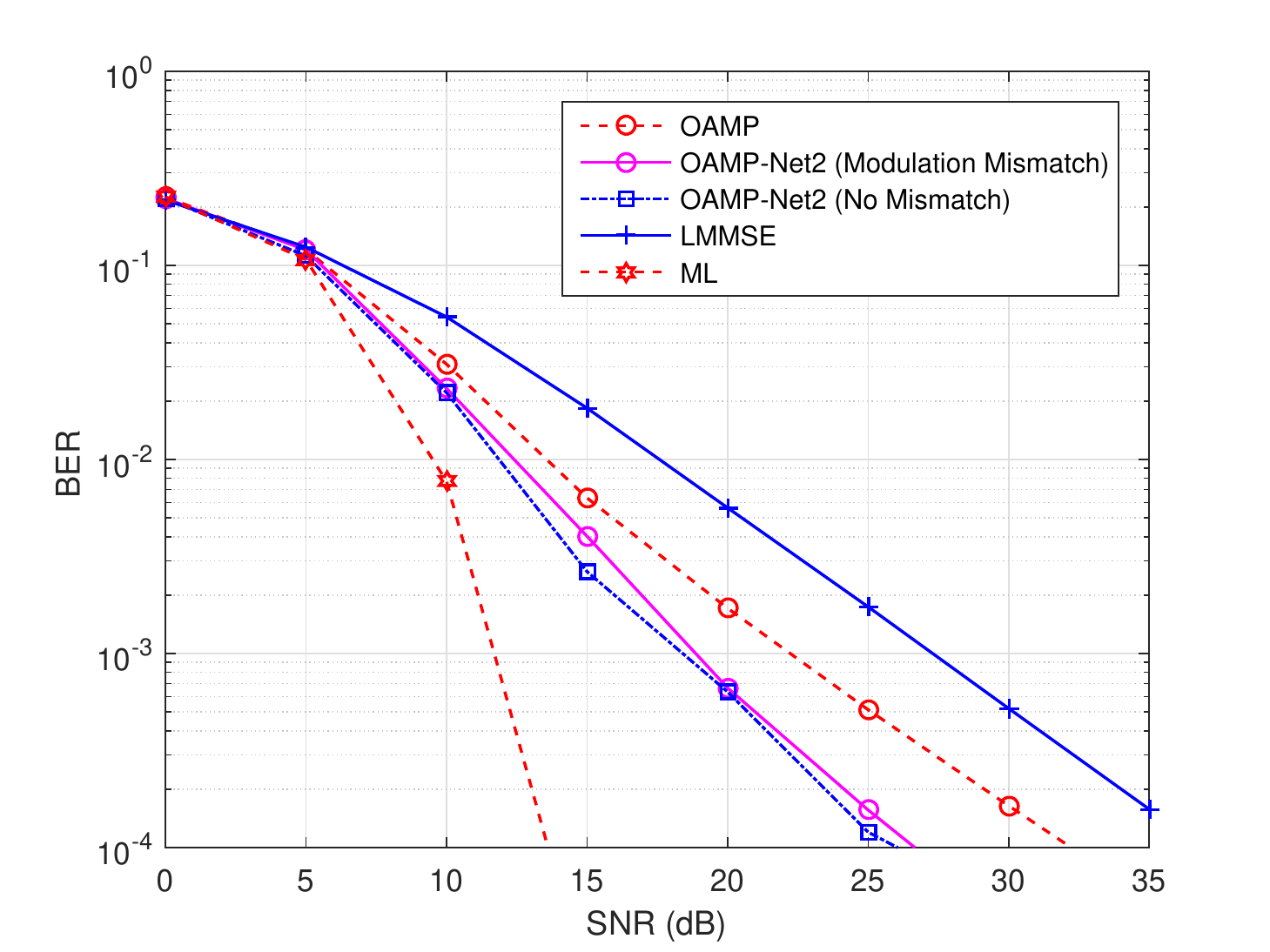}}
  \centerline{(b) Modulation symbol mismatch}
\end{minipage}\label{Robustness2b}
\caption{.~~BERs performance comparisons of the OAMP and OAMP-Net2 with the MIMO configuration and modulation symbol mismatches.}
\label{Robustness_2}
\end{figure}

Fig.\,\ref{Robustness_2} presents the BER performance of OAMP and OAMP-Net2 with the MIMO configuration and modulation symbol mismatches. We train the OAMP-Net2 with $16$-QAM symbol in $8\times8$ MIMO system and test the robustness of the trained parameters. Fig.\,10(a) exhibits the numerical results of the OAMP-Net2 with MIMO configuration mismatch, where the network is tested in $4\times4$ systems with $16$-QAM. Although the network is employed in different MIMO configurations, the OAMP-Net2 still outperforms the LMMSE and OAMP detectors and has little performance loss. The robustness demonstrates the OAMP-Net2 is flexible to different MIMO configuration. Furthermore, Fig.\,10(b) shows the performance of OAMP-Net2 with modulation symbol mismatch, where the network is tested in $8\times8$ MIMO system with QPSK symbol. Only $0.4$ dB performance loss is incurred owing to modulation symbol mismatch if we target for BER=$10^{-2}$, which demonstrates the OAMP-Net2 is robust to modulation symbol mismatch. The OAMP-Net2 detector contains a linear and a nonlinear estimators. The linear estimator is related to gradient descent algorithm and its convergence behavior and performance are determined by appropriate step-size $\gamma_{t}$ moving to the search point. As the linear estimator is independent of the modulation symbols, the learnable parameter $\gamma_{t}$ shows strong robustness to modulation symbol mismatches. Therefore, the trained network can be utilized directly in different MIMO configurations with different modulation symbols.

\section{Conclusion}\label{con}
We have developed a novel model-driven DL network for MIMO detection, named OAMP-Net2,  which is obtained by unfolding the OAMP detection algorithm. The OAMP-Net2 detector inherits the superiority of the Bayes-optimal signal recovery algorithm and DL technique, and thus presents excellent performance. The network are easy and fast to train because only few adjustable parameters are required to be optimized. To handle imperfect CSI, an OAMP-Net2-based JCESD is proposed, where the detector takes channel estimation error and channel statistics into consideration while channel estimation is refined by
detected data and considers the detection error.
Simulation results demonstrate 
that significant performance gain can be obtained by learning corresponding optimal parameters from the data to improve the detector and compensate for the channel estimation error. Furthermore, the OAMP-Net2 exhibits strong robustness to various mismatches.
\section{Acknowledgement}\label{ACK}
The authors would like to thank Dr. Junjie Ma for discussing the structure of the OAMP algorithm.

\section*{Appendix A: derivation for Covariance matrices}\label{AP2}
In order to derive the covariance matrix $\hat{\mathbf{V}}_{\rmd}[n]$ of the equivalent noise $\hat{\mathbf{n}}_{\rmd}[n]$ in signal detection stage, we should evaluate the statistical properties of the $\mathbf{z}_{\rmd}[n]=\bm{\Delta}\bH\mathbf{x}_{\rmd}[n]$ for $n = 1,\ldots,\Nd $, which can be expressed as
\begin{equation}\label{Covzd}
  \mathrm{Cov}[\mathbf{z}_{\rmd}[n]]=\mathtt{E}[\mathbf{z}_{\rmd}[n](\mathbf{z}_{\rmd}[n])^{H}].
\end{equation}
As the covariance matrix $\mathrm{Cov}[\mathbf{z}_{\rmd}[n]]$  for each symbol index $n = 1,\ldots,\Nd $ is same and the data symbol is independent  between different antennas, we omit the index $n$  and have
\begin{equation}
\mathtt{E}[z_{\rmd,i}z_{\rmd,k}^{H}] = \left\{
\begin{aligned}
 & 0, \quad \forall i \neq k \\
& \sum_{j=1}^{\Nt}\sigma_{\Delta h_{i,j}}^{2}, \quad i = k
\end{aligned}
\right.
\end{equation}
where $z_{\rmd,i}=\sum_{j=1}^{\Nt}\Delta h_{i,j}x_{\rmd,j}$ is the $i$-th element of the $\mathbf{z}_{\rmd}$, $\Delta h_{i,j}$ is the $(i,j)$-th element of the channel estimation error matrix $\bm{\Delta}\bH$, and $x_{\rmd,j}$ is $j$-th element of the $\bx_{\rmd}$. The $\sigma_{\Delta h_{i,j}}^{2}$ is the variance of $(i,j)$-th element in channel estimation error matrix $\bm{\Delta}\bH$ which can be obtained from $\bR_{\bm{\Delta}\bh_{\rmp}}$ in (\ref{LMMSE_error}) or $\bR_{\bm{\Delta}\bh}$ in (\ref{LMMSE_error2}). By considering the contribution of original additive noise, the covariance matrix $\hat{\mathbf{V}}_{\mathrm{est}}$ can be obtained as
\begin{equation}\label{covvd}
  \hat{\mathbf{V}}_{\mathrm{est}} = \mathrm{diag} \left(\sum_{j=1}^{\Nt}\sigma_{\Delta h_{i,j}}^{2}+\sigma^{2},\ldots,\sum_{j=1}^{\Nt}\sigma_{\Delta h_{i,j}}^{2}+\sigma^{2} \right).
\end{equation}
The covariance matrix $\hat{\mathbf{V}}_{\mathrm{est}}$ will be utilized in the OAMP-Net2 detector as the covariance matrix of the equivalent noise $\bR_{\hat{\bn}_{\rmd}\hat{\bn}_{\rmd}}$.

Next, we compute the covariance matrix $\hat{\bV}_{\rmp}[n]$ of the equivalent noise $\hat{\bn}_{\rmp}[n]$ in data-aided channel estimation stage using similar approach. For each time index $n$, We denote $\bz_{n}=\mathbf{H}\bee_{n}$ and $z_{i,n} = \sum_{j=1}^{\Nt}h_{i,j}e_{j,n}$, where $\bee_{n}$ is the $n$-th column of the signal detection error matrix $\bE_{\rmd}$. In a similar way, we have
\begin{equation}
\mathtt{E}[z_{i,n}z_{k,n}^{H}] = \left\{
\begin{aligned}
& 0, \quad \forall i \neq k \\
& \sum_{j=1}^{\Nt}\sigma_{e_{j,n}}^{2}/\Nr, \quad  i = k
\end{aligned}
\right.
\end{equation}
because $h_{i,j} \sim \mathcal{N}_{\mathbb{C}}(0,1/\Nr)$ and $\sigma_{e_{j,n}}^{2}$ is the variance for $j$-th element of the $\bee_{n}$.
Therefore, the covariance matrix $\hat{\mathbf{V}}_{\rmp}[n]$ is given by
\begin{equation}\label{eqvp}
  \hat{\mathbf{V}}_{\rmp}[n] = \left(\sum_{j=1}^{\Nt}\sigma_{e_{j,n}}^{2}+\sigma^{2}\right)\mathbf{I}_{\Nt}.
\end{equation}
The covariance matrix $\hat{\mathbf{V}}_{\rmp}[n]$ will be utilized in channel estimator to evaluate the covariance matrix of the equivalent noise $\bR_{\bn\bn}$ where $\bn=\mathrm{vec}(\hat{\bN})$. By considering the different covariance matrix of the actual pilot and additional pilot, we have
\begin{align}\label{finalSVD}
\bR_{\bn \bn} = \left[\begin{array}{c c c}
 \sigma^{2}\bI_{\Np\Nr}  \\
   & \hat{\mathbf{V}}_{\mathrm{det}}
\end{array}\right]
\end{align}
where $\sigma^{2}\bI_{\Np\Nr}$ denotes the covariance matrix of the noise $\mathrm{vec}(\bN_{\rmp})$ for actual pilot $\bX_{\rmp}$ while
\begin{align}\label{finalSVD}
 \hat{\mathbf{V}}_{\mathrm{det}} = \left[\begin{array}{c c c}
 \hat{\mathbf{V}}_{\rmp}[1]  \\
 &\ddots   \\
 &&\hat{\mathbf{V}}_{\rmp}[\Nd]
\end{array}\right]
\end{align}
denotes the covariance matrix of the equivalent noise $\mathrm{vec}(\hat{\bN}_{\rmp})$ for the additional pilot $\hat{\bX}_{\rmd}$.

\section*{Appendix B: derivation for variance estimators}\label{AP1}
 To derive the expressions for the error variance estimators $v_{t}^{2}$ and $\tau_{t}^{2}$ in OAMP-Net2, we use similar method to \cite{TISTA}. We should indicate that $v_{t}^{2}$ and $\tau_{t}^{2}$ in OAMP-detector are based on the following two assumptions about error vectors $\mathbf{p}_{t}$ and $\mathbf{q}_{t}$ \cite{OAMP}.

\emph{Assumption 1:} $\mathbf{p}_{t}$ consists of i.i.d. zero-mean Gaussian entries independent of $\mathbf{x}_{\rmd}$.

\emph{Assumption 2:} $\mathbf{q}_{t}$ consists of i.i.d. zero-mean Gaussian entries independent of $\mathbf{H}$ and $\hat{\mathbf{n}}_{\rmd}$.

 From the similar viewpoint, we obtain the error variance estimators $v_{t}^{2}$ and $\tau_{t}^{2}$  ($\ref{eqlv}$) and ($\ref{eqlt}$ in OAMP-Net2
 using \emph{Assumption 1} and \emph{2}, and considering the effect of the learnable variables $\gamma_{t}$ and $\theta_{t}$. For convenience, we will use $\bH$, $\hat{\bx}_{t}$, and $\bn$ to refer $\hat{\bH}$, $\hat{\bx}_{\rmd,t}$, and $\hat{\bn}_{\rmd}$, respectively. Based on \emph{assumption 2}, $\mathbf{q}_{t}$ consists of i.i.d. zero-mean Gaussian entries which is independent of $\mathbf{H}$ and $\mathbf{n}$, which means that
 \begin{equation}\label{eqA2}
\mathbb{E}[(\hat{\mathbf{x}}_{t}-\mathbf{x})^{H}\mathbf{H}^{H}\mathbf{n}]=0.
\end{equation}
 Therefore, The error variance estimator ${v}_{t}^{2}$ is given by,
\begin{align}\label{eqv2}
   {v}_{t}^{2} & =\frac{\mathtt{E}[\|\mathbf{q}_{t}\|^{2}_{2}]}{\Nt} = \frac{1}{\Nt}\frac{\mathrm{tr}(\bH^{H}\bH)
  \mathtt{E}[\|\mathbf{q}_{t}\|^{2}_{2}]}{\mathrm{tr}(\mathbf{H}^{H}\mathbf{H})} \nonumber \\
& =\frac{\mathtt{E}[(\mathbf{x}-\hat{\mathbf{x}}_{t})^{H}\mathbf{H}^{H}\mathbf{H}(\mathbf{x}-\hat{\mathbf{x}}_{t})]}{\mathrm{tr}(\mathbf{H}^{H}\mathbf{H})} \nonumber \\
& \overset{a}{=}
\frac{\mathtt{E}[\left(\mathbf{H}(\mathbf{x}-\hat{\mathbf{x}}_{t})\right)^{H}\mathbf{H}(\mathbf{x}-\hat{\mathbf{x}}_{t})+\left(\mathbf{H}(\mathbf{x}-\hat{\mathbf{x}}_{t})\right)^{H}\mathbf{n}]}
{\mathrm{tr}(\mathbf{H}^{H}\mathbf{H})} \nonumber \\
& = \frac{\mathtt{E}[\|\mathbf{H}(\mathbf{x}-\hat{\mathbf{x}}_{t})+\mathbf{n}\|^{2}_{2}]-\mathtt{E}[\mathbf{n}^{H}\mathbf{n}]}
{\mathrm{tr}(\mathbf{H}^{H}\mathbf{H})} \nonumber \\
& = \frac{\mathtt{E}[\|  \mathbf{H}\mathbf{x}+\mathbf{n} - \mathbf{H}\hat{\mathbf{x}}_{t}  \|^{2}_{2}]- \Nr\sigma^{2}}
{\mathrm{tr}(\mathbf{H}^{H}\mathbf{H})}   \nonumber \\
& = \frac{\mathtt{E}[\|  \mathbf{y} - \mathbf{H}\hat{\mathbf{x}}_{t}  \|^{2}_{2}]- \Nr\sigma^{2}}
{\mathrm{tr}(\mathbf{H}^{H}\mathbf{H})},
\end{align}
where $\overset{a}{=}$ is obtained by using the equation ($\ref{eqA2}$). The final expression for ${v}_{t}^{2}$ is obtained by ($\ref{eqv2}$).

Next, we derive the expression for $\tau_{t}^{2}$. The error vector $\mathbf{r}_{t}-\mathbf{x}$ can be rewritten as
\begin{align}\label{eqr2}
   \mathbf{r}_{t}-\mathbf{x} & = \hat{\mathbf{x}}_{t}+\gamma_{t}\mathbf{W}_{t}(\mathbf{y}-\mathbf{H}\hat{\mathbf{x}}_{t})-\mathbf{x} \nonumber \\
& = \hat{\mathbf{x}}_{t} + \gamma_{t}(\mathbf{H}\mathbf{x}+\mathbf{n})-\gamma_{t}\mathbf{W}_{t}\mathbf{H}\hat{\mathbf{x}}_{t}-\mathbf{x} \nonumber \\
& = (\mathbf{I}-\gamma_{t}\mathbf{W}_{t})(\hat{\mathbf{x}}_{t}-\mathbf{x}) + \gamma_{t}\mathbf{W}_{t}\mathbf{n}.
\end{align}
Then, for the error variance estimator $\tau_{t}^{2}$, we have
\begin{align}\label{eqtau2}
\tau_{t}^{2} & = \frac{\mathtt{E}[\|\mathbf{q}_{t}\|^{2}_{2}]}{\Nt} = \frac{1}{\Nt}\mathtt{E}[\|(\mathbf{I}-\gamma_{t}\mathbf{W}_{t}\mathbf{H})
(\hat{\mathbf{x}}_{t}-\mathbf{x}) +  \gamma_{t}\mathbf{W}_{t}\mathbf{n}\|^{2}_{2}] \nonumber \\
& =  \frac{1}{\Nt} \mathtt{E}[(\hat{\mathbf{x}}_{t}-\mathbf{x})^{H}(\mathbf{I}-\gamma_{t}\mathbf{W}_{t}\mathbf{H})(\mathbf{I}-\gamma_{t}\mathbf{W}_{t}\mathbf{H})^{H}
(\hat{\mathbf{x}}_{t}-\mathbf{x})] \nonumber \\
& + \frac{\gamma_{t}^{2}}{\Nt}\mathtt{E}[\mathbf{n}^{H}\mathbf{W}_{t}^{H}\mathbf{W}_{t}\mathbf{n}]
+\frac{2\gamma_{t}}{\Nt}\mathtt{E}[(\hat{\mathbf{x}}_{t}-\mathbf{x})^{H}(\mathbf{I}-\gamma_{t}\mathbf{W}_{t}\mathbf{H})^{H}\mathbf{W}_{t}\mathbf{n}] \nonumber \\
& = \frac{1}{\Nt}\mathrm{tr}\left((\mathbf{I}-\gamma_{t}\mathbf{W}_{t}\mathbf{H})(\mathbf{I}-\gamma_{t}\mathbf{W}_{t}\mathbf{H})^{H}\right){v}_{t}^{2} \nonumber \\
& + \frac{\gamma_{t}^{2}}{\Nt}\mathrm{tr}(\mathbf{W}_{t}{\mathbf{W}_{t}}^{H})\sigma^{2}
+ \frac{2(\gamma_{t}-\gamma_{t}^{2})}{\Nt} \mathtt{E}[(\hat{\mathbf{x}}_{t}-\mathbf{x})^{H}\mathbf{H}^{H}\mathbf{n}].
\end{align}
The last term in the ($\ref{eqtau2}$) vanishes as the $\mathtt{E}[(\hat{\mathbf{x}}_{t}-\mathbf{x})^{H}\mathbf{H}^{H}\mathbf{n}]=0$. Therefore, the error variance estimator $\tau_{t}^{2}$ is given by
\begin{align}\label{eqtau2re}
  \tau_{t}^{2} &= \frac{1}{\Nt}\mathrm{tr}\left((\mathbf{I}-\gamma_{t}\mathbf{W}_{t}\mathbf{H})(\mathbf{I}-\gamma_{t}\mathbf{W}_{t}\mathbf{H})^{H}\right){v}_{t}^{2} \nonumber \\
  & + \frac{\gamma_{t}^{2}}{\Nt}\mathrm{tr}(\mathbf{W}_{t}{\mathbf{W}_{t}}^{H})\sigma^{2}.
\end{align}
Then, we replace the parameter $\gamma_{t}$ with $\theta_{t}$ in ($\ref{eqtau2re}$) as the simulation results show that two parameters ($\gamma_{t}$, $\theta_{t}$) can  preferably regulate the error variance estimator ${v}_{t}^{2}$ and $\tau_{t}^{2}$. The superior performance is attributed to more parameters, which can incorporate more side information from the data. Therefore, the final expression for $\tau_{t}^{2}$ is obtained by
\begin{equation}\label{eqltt}
  \tau^{2}_{t}=\frac{1}{\Nt}\mathrm{tr}(\mathbf{C}_{t}\mathbf{C}_{t}^{H})v_{t}^{2}+\frac{\theta_{t}^{2}\sigma^{2}}{\Nt}\mathrm{tr}(\mathbf{W}_{t}\mathbf{W}_{t}^{H}),
\end{equation}
where $\mathbf{C}_{t}=\mathbf{I}-\theta_{t}\mathbf{W}_{t}\mathbf{H}$.


\end{document}